\documentclass[12pt]{article}
\usepackage{amssymb}
\usepackage{amsmath}
\usepackage[dvips]{color}
\definecolor{g3}{rgb}{0.8,0.8,0.8}
\def\hybrid{\topmargin 0pt      \oddsidemargin 0pt
        \headheight 0pt \headsep 0pt
        \textwidth 16.5cm
        \textheight 23cm
        \voffset=-1cm
        \hoffset=0.4cm
        \marginparwidth 0.0in
        \parskip 5pt plus 1pt   \jot = 1.5ex}
\catcode`\@=11
\def\marginnote#1{}

\newcount\hour
\newcount\minute
\newtoks\amorpm
\hour=\time\divide\hour by60 \minute=\time{\multiply\hour by60
\global\advance\minute by-\hour}
\edef\standardtime{{\ifnum\hour<12 \global\amorpm={am}%
        \else\global\amorpm={pm}\advance\hour by-12 \fi
        \ifnum\hour=0 \hour=12 \fi
      \number\hour:\ifnum\minute<10 0\fi\number\minute\the\amorpm}}
\edef\militarytime{\number\hour:\ifnum\minute<10
0\fi\number\minute}

\def\draftlabel#1{{\@bsphack\if@filesw {\let\thepage\relax
   \xdef\@gtempa{\write\@auxout{\string
      \newlabel{#1}{{\@currentlabel}{\thepage}}}}}\@gtempa
   \if@nobreak \ifvmode\nobreak\fi\fi\fi\@esphack}
        \gdef\@eqnlabel{#1}}
\def\@eqnlabel{}
\def\@vacuum{}
\def\draftmarginnote#1{\marginpar{\raggedright\scriptsize\tt#1}}

\def\draft{\oddsidemargin -0.1truein
        \def\@oddfoot{\sl preliminary draft \hfil
        \rm\thepage\hfil\sl\today\quad\militarytime}
        \let\@evenfoot\@oddfoot \overfullrule 3pt
        \let\label=\draftlabel
        \let\marginnote=\draftmarginnote
\def\@eqnnum{{\rm (\theequation)}
\rlap{\kern\marginparsep\tt\@eqnlabel}%
\global\let\@eqnlabel\@vacuum}  }
\def\emline#1#2#3#4#5#6{%
       \put(#1,#2){\special{em:moveto}}%
       \put(#4,#5){\special{em:lineto}}}

\renewcommand{\AA}{{\mathbb{A}}}
\newcommand{\RR}{{\mathbb{R}}}
\newcommand{\CC}{{\mathbb{C}}}

\newcommand{\ZZ}{{\mathbb{Z}}}            

\newfont{\Bbbb}{msbm7 scaled 1\@ptsize00}
\newcommand{\zs}{\raise-1pt\hbox{$\mbox{\Bbbb Z}$}}
\newcommand{\rs}{\hbox{$\mbox{\Bbbb R}$}}

\font\teneufm=cmmib10 scaled 1\@ptsize00 \font\seveneufm=cmmib7
scaled 1\@ptsize00
\font\fiveeufm=cmmib5  
\def\bfit#1{{\textfont1=\teneufm\scriptfont1=\seveneufm
\scriptscriptfont1=\fiveeufm \mathchoice{
\hbox{$\mathsurround=0pt\displaystyle#1$}}
{\mathsurround=0pt\hbox{$\textstyle#1$}}
{\hbox{$\mathsurround=0pt\scriptstyle#1$}}
{\hbox{$\mathsurround=0pt\scriptscriptstyle#1$}}}}

\def\numberbysection{\@addtoreset{equation}{section}
        \def\theequation{\thesection.\arabic{equation}}}
\numberbysection

\renewcommand{\theequation}{\thesection.\arabic{equation}}
\newcommand{\l@qq}[2]{\addvspace{2em}
 \hbox to\textwidth{\hspace{1em}\bf #1 \dotfill #2}}

\newcounter{app}

\def\app{\setcounter{equation}{0}
\def\theequation{\Alph{app}.\arabic{equation}}\par
   \addvspace{10ex}
   \@afterindentfalse
  \secdef\@app\@dapp}
\newcommand\@app{\@startsection {app}{1}{-0.3ex}%
                             {-3.5ex \@plus -1ex \@minus -.2ex}%
                                   {2.3ex \@plus.2ex}%
                                   {\normalfont\Large\bf}}

\def\@dapp#1{%
{\parindent \z@ \raggedright  \bf #1}\par\nobreak}
\def\l@app#1#2{\ifnum \c@tocdepth >\z@
    \addpenalty\@secpenalty
    \addvspace{1.0em \@plus\p@}%
    \setlength\@tempdima{1.5em}%
    \begingroup
      \parindent \z@ \rightskip \@pnumwidth
      \parfillskip -\@pnumwidth
      \leavevmode \bfseries
      \advance\leftskip\@tempdima
      \hskip -\leftskip
      #1\nobreak\hfil \nobreak\hb@xt@\@pnumwidth{\hss #2}\par
    \endgroup\fi}
\newcounter{sapp}[app]

\def\sapp{\def\theequation{\Alph{app}.\arabic{equation}}\par
   \@afterindentfalse
  \secdef\@sapp\@dsapp}
\newcommand\@sapp{\@startsection{sapp}{2}{\z@}%
                           {-3.25ex\@plus -1ex \@minus -.2ex}%
                           {1.5ex \@plus .2ex}%
                              {\normalfont\large\bfseries}}

\def\@dsapp#1{%
{\parindent \z@ \raggedright  \bf #1}\par\nobreak}

\newcommand{\l@sapp}{\@dottedtocline{2}{1.4em}{2.5em}}

\def\titlepage{\@restonecolfalse\if@twocolumn\@restonecoltrue\onecolumn
     \else \newpage \fi \thispagestyle{empty}\c@page\z@
        \def\thefootnote{\fnsymbol{footnote}} }

\def\endtitlepage{\if@restonecol\twocolumn \else  \fi
        \def\thefootnote{\arabic{footnote}}
        \setcounter{footnote}{0}}  
\relax

\hybrid

\newtoks\@stequation

\def\subequations{\refstepcounter{equation}%
  \edef\@savedequation{\the\c@equation}%
  \@stequation=\expandafter{\theequation}
  \edef\@savedtheequation{\the\@stequation}
  \edef\oldtheequation{\theequation}%
  \setcounter{equation}{0}%
  \def\theequation{\oldtheequation\alph{equation}}}

\def\endsubequations{%
  \setcounter{equation}{\@savedequation}%
  \@stequation=\expandafter{\@savedtheequation}%
  \edef\theequation{\the\@stequation}%
  \global\@ignoretrue}

\makeatletter
\newdimen\normalarrayskip            
\newdimen\minarrayskip               
\normalarrayskip\baselineskip \minarrayskip\jot
\newif\ifold             \oldtrue            \def\new{\oldfalse}
\def\arraymode{\ifold\relax\else\displaystyle\fi}
\def\eqnumphantom{\phantom{(\theequation)}} 
\def\@arrayskip{\ifold\baselineskip\z@\lineskip\z@
     \else
     \baselineskip\minarrayskip\lineskip1\baselineskip\fi}

\def\@arrayclassz{\ifcase \@lastchclass \@acolampacol \or
\@ampacol \or \or \or \@addamp \or
   \@acolampacol \or \@firstampfalse \@acol \fi
\edef\@preamble{\@preamble
  \ifcase \@chnum
     \hfil$\relax\arraymode\@sharp$\hfil
     \or $\relax\arraymode\@sharp$\hfil
     \or \hfil$\relax\arraymode\@sharp$\fi}}

\def\@array[#1]#2{\setbox\@arstrutbox=\hbox{\vrule
     height\arraystretch \ht\strutbox
     depth\arraystretch \dp\strutbox
width\z@}\@mkpream{#2}\edef\@preamble{\halign \noexpand\@halignto
\bgroup \tabskip\z@ \@arstrut \@preamble \tabskip\z@ \cr}%
\let\@startpbox\@@startpbox \let\@endpbox\@@endpbox
  \if #1t\vtop \else \if#1b\vbox \else \vcenter \fi\fi
  \bgroup \let\par\relax
  \let\@sharp##\let\protect\relax
  \@arrayskip\@preamble}
\def\eqnarray{\stepcounter{equation}%
              \let\@currentlabel=\theequation
              \global\@eqnswtrue
              \global\@eqcnt\z@
              \tabskip\@centering              
              \let\\=\@eqncr
              $$%
            \halign to \displaywidth  \bgroup
             \eqnumphantom \@eqnsel
      \hskip\@centering                               
    $\displaystyle  \tabskip\z@ {##}$%
    &\global\@eqcnt\@ne \hskip 2\arraycolsep
         $ \displaystyle  \arraymode{##}$\hfil
    &\global\@eqcnt\tw@ \hskip 2\arraycolsep
         $\displaystyle\tabskip\z@{##}$\hfil
         \tabskip\@centering
    &{##}\tabskip\z@\cr}
\makeatother
\newtheorem{te}{Theorem}[section]
\newtheorem{de}{Definition}[section]
\newtheorem{prop}{Proposition}[section]           
\newtheorem{cor}{Corollary}[section]
\newtheorem{lem}{Lemma}[section]

\newtheorem{rem}{Remark}[section]

\def\bea{\begin{eqnarray}}
\def\eea{\end{eqnarray}}

\def\beq{\begin{equation}}
\def\eeq{\end{equation}}
\def\be{\bea\new\begin{array}{c}}  
\def\ee{\end{array}\eea}           
\def\bse{\begin{subequations}}                
\def\ese{\end{subequations}}

\def\square{\hfill{\vrule height6pt width6pt            
depth1pt} \break \vspace{.01cm}}                        
\def\d{\partial}

\def\tr{\triangleright}                                
\def\tl{\triangleleft}                                 

\def\jo{\mathrel{\mkern-4mu}}
\def\sem{\mathsurround=0pt
\mathrel{\raise1.4pt\hbox{$\scriptscriptstyle>$}}\jo\mathrel\tl}
\def\mes{\mathsurround=0pt
{\mathrel\tr\jo\mathrel{\raise1.4pt\hbox{$\scriptscriptstyle
<$}}}}
\def\bal{{\bfit\alpha}}

\def\bgamma{{\bfit\gamma}}

\def\blambda{{\bfit\lambda}}
\def\brho{{\bfit\rho}}
\def\bo{{\bfit\omega}}
\def\al{\alpha}
\def\la{\lambda}

\def\e{\epsilon}
\def\<{\langle}
\def\>{\rangle}
\def\ov{\overline}
\def\wt{\widetilde}
\def\wh{\widehat}

\def\N{\scriptscriptstyle N}
\def\zen{g}
\begin{document}

\title{
\hfill {\normalsize ITEP-TH-8/01}\\[10mm]
\bf Unitary representations of
$U_{q}(\mathfrak{sl}(2,\RR))$, the modular double, and the
multiparticle $q$-deformed Toda chains}

\author{S. Kharchev\thanks{Institute of Theoretical and Experimental
Physics, Moscow, Russia}%
\and D. Lebedev\thanks{Institute of Theoretical and Experimental
Physics, Moscow, Russia} %
\and M. Semenov-Tian-Shansky\thanks{Universit\'e de %
Bourgogne, Dijon, France, and Steklov Math. Institute,
St.Petersburg, Russia }}
\date{\phantom{.}}
\maketitle

\begin{abstract}
The paper deals with the analytic theory of the quantum
$q$-deformed Toda chains; the technique used combines the methods
of representation theory and the Quantum Inverse Scattering Method.
The key phenomenon which is under scrutiny is the role of the
modular duality concept (first discovered by L.Faddeev) in the
representation theory of noncompact semisimple quantum groups.
Explicit formulae for the Whittaker vectors are  presented in
terms of the double sine functions and the wave functions of the
$N$-particle $q$-deformed open Toda chain are given as a multiple
integral of the Mellin-Barnes type. For the periodic chain the
two dual Baxter equations are derived.

\end{abstract}

\section*{Preface}
In the late seventies B. Kostant  \cite{Ko}  has discovered a
fascinating link between the representation   theory of
non-compact semisimple Lie groups and the  quantum Toda chain.
Let $G$ be a real
split semisimple Lie group, $B=MAN$ its minimal Borel subgroup, let
$N$ and $V=\bar{N}$ be the corresponding opposite unipotent subgroups.
Let $\chi _{N},\chi _{V}$ be nondegenerate unitary characters of $N$
and $V$, respectively. Let ${\cal H}_{T}$ be the space of smooth
functions on $G$ which satisfy the functional equation
$$
\varphi(vxn)=\chi _{V}(v)\overline{\chi _{N}(n)}\,\varphi(x),
\ \ \ \ \ v\in V, n\in N.
$$
A function $\varphi \in {\cal H}_{T}$ is uniquely determined by its
restriction to $A\subset G$. Obviously, ${\cal H}_{T}$ is invariant
under the action of the center of the universal enveloping algebra
$Z\subset U(\mathfrak{g})$; hence, any Casimir
operator $C\in Z$  gives rise to a differential operator acting in $%
C^{\infty }(A).$ When $C$ is the quadratic Casimir, this is
precisely the Toda Hamiltonian; other Casimirs provide a complete
set of quantum integrals of motion.

This observation reduces the spectral theory of Toda chain to
the representation theory of semisimple Lie groups. The joint
eigenfunctions of the quantum Toda Hamiltonians are the so called
generalized Whittaker functions. The theory of Whittaker functions
has been extensively studied in the 60's and 70's
\cite{Jac}, \cite{Sch}, \cite{Ha}; it displays
deep parallels with the celebrated Harish-Chandra theory of
spherical functions \cite{HC} and depends on a profound study of the
principal series representations  \cite{STS-toda}.

The group theoretic approach based on representation theory of
finite-dimensional semisimple groups is matched by a more
sophisticated technique of the Quantum Inverse Scattering Method
\cite{QISM}. The treatment of the Toda chain by means of QISM is
based on a 2$\times $2 matrix first order difference Lax operator
for the Toda lattice. (In order to understand its relation to the
$n\times n$ Lax representation which is implicit in Kostant's
approach recall that the Lax matrix is a tridiagonal Jacobi matrix
which defines a three-term recurrence relation and hence may be
regarded as a second order scalar difference operator). While the
use of the lattice Lax representation restricts
generality: we have to assume that ${\mathfrak g}=\mathfrak{sl}(n)$
\footnote{
The treatment of other classical Lie algebras is also possible;
for that end, one needs to use lattice Lax pairs
with boundary conditions introduced by Sklyanin \cite{Skl3}.
In the present note we shall not deal with this generalization
and assume that $\mathfrak{g}=\mathfrak{sl}(n)$
},
it allows to bring into play the powerful machinery of quantum
$R$-matrices (and hence eventually of {\em infinite dimensional}\/
quantum groups). Recently the first two authors have established an
explicit connection of  the QISM-based approach to the quantum Toda
chain to the theory of Whittaker functions \cite{KL2}. The
technique of QISM yields new explicit formulae for the Whittaker
functions which, to the best of our knowledge, were not known in
the elementary representation theory.

It looks  rather natural to generalize this   approach   to the
$q$-deformed case. The use of lattice Lax representation makes the
procedure rather straightforward: one simply has to replace the
rational $R$-matrix with the trigonometric one (we shall see,
however, that this generalization includes a number of nontrivial
points). On the other hand, the very definitions of `noncompact
quantum groups' which one needs to proceed with the $q$-deformed
version of the Kostant approach are by no means obvious. It is the
interplay of the  explicit formulae based on QISM and of their
not-yet-defined counterparts coming from the representation theory of
noncompact finite-dimensional quantum groups that makes the entire
game very exciting. Our preliminary results suggest that the correct
treatment of the problem requires a very significant change in the
entire framework of the representation theory of
$U_{q}(\mathfrak{g})$; the crucial role is played by the `modular
dual' of $U_{q}(\mathfrak{g})$ and the {\em modular double}
$U_{q}(\mathfrak{g})\otimes U_{\wt q}\,(\mathfrak{g})$ which was
introduced recently by Faddeev \cite{Fad2} \footnote{ The definition
of the modular double was coined out by Faddeev in the special case
${\mathfrak g}=\mathfrak{sl}(2)$; as pointed out to the authors by
B.~Feigin, it is most likely that for general semisimple Lie algebras
the modular dual of $U_{q}(\mathfrak{g})$ is
$U_{\wt q}\,(\check{\mathfrak{g}})$, where $\check{\mathfrak{g}}$ is
the Langlands dual of $\mathfrak{g}$. }. Among other things, this new
point of view  leads to new possibilities in the choice of real forms
of the relevant algebras: it is the real form of the modular double
$U_{q}(\mathfrak{g})\otimes U_{\wt q}\,(\mathfrak{g})$ which really
matters. One nontrivial possibility for the choice of the real form
has been recently pointed out by Faddeev, Kashaev and Volkov
\cite{FKV} in their study of the quantum Liouville theory; it is very
encouraging that the same real form naturally arises in the study of
the $q$-deformed Toda chain.

Analytical aspects of the theory bring into play the double gamma
and double sine functions of Barnes \cite{Ba1, Ba2, Ba3, Shin, Kur},
or the closely related quantum dilogarithms  \cite{FK},  which
replace the ordinary gamma functions in the formulae for both the
Harish-Chandra $c$-functions and the Whittaker functions. We believe
that the implications of these constructions for the
representation theory are probably more interesting than the
$q$-deformed Toda model itself (commonly known as the relativistic
Toda chain \cite{Rui}).

Our strategy in the present paper is as follows. In Section 1
we shall start with the elementary representation theory of the
algebra $U_{q}(\mathfrak{sl}(2,\RR))$. Section 2 deals with the
theory of Whittaker vectors and Whittaker functions for
the modular double $U_q(\mathfrak{sl}(2,\RR))\otimes
U_{\,\wt q}(\mathfrak{sl}(2,\RR))$ and with the 2-particle
$q$-deformed open Toda chain. We obtain explicit formulae for
the Whittaker vectors in terms of the double sine functions
and derive the integral representations for solutions to
one-parameter family of two-particle relativistic Toda chains in
the framework of representation theory; all these solutions enjoy
with the dual symmetry.
Generalization to the $N$-particle case is described in
Section 3; using the QISM approach, we derive an appropriate
solution to the spectral problem for the open $N$-particle chain
in the form of multiple integral of the Mellin-Barnes type with
the natural deformation of the usual gamma functions to the double
sine functions. It is shown that the solution for the $N$-periodic
chain is represented as a generalized Fourier transform
of $N-1$-particle open wave function with the kernel satisfying
two mutually dual Baxter equations. Finally, in Appendix we list
the essential analytic properties of the double sine functions.

\section{Representations of principal series of
$U_{q}(\mathfrak{sl}(2,\RR))$ and the Modular Double}

In this section we shall discuss the representations of
$U_{q}(\mathfrak{sl}(2,\RR))$ which may be regarded as deformations
of the principal series representations of $SL(2,{\RR})$. As pointed
out by Faddeev \cite{Fad2}, these representations
possess a remarkable  duality which is similar to the modular
duality for noncommutative tori discovered by Rieffel \cite{R}.
We start with the algebraic definition of
$U_{q}(\mathfrak{sl}(2,\CC))$ (see, for example, \cite{ChP}).
It is generated by elements $K^{\pm 1},E,F$ subject to the relations
\be\label{uqsl2}
KE=q^2EK\ \ \ \ \ \ KF=q^{-2}FK,\\
EF-FE=\frac{K-K^{-1}}{q-q^{-1}},
\ee
where
\be
q=e^{\pi i \tau}, \ \ \ \ \tau\in\CC
\ee
The bialgebra structure on $U_{q}(\mathfrak{sl}(2,\CC))$
is given by the coproduct
\footnote{
The coalgebraic structure of $U_{q}(\mathfrak{sl}(2,\CC))$
is not used in the present paper.
}
\be\label{hopf}
\Delta K = K\otimes K,\\
\Delta E = E\otimes 1+K\otimes E,\\
\Delta F = 1\otimes F+F\otimes K^{-1}.
\ee
The center of $U_{q}(\mathfrak{sl}(2,\CC))$ is generated by
the Casimir element
\be\label{casimir}
C_2=qK+q^{-1}K^{-1}+(q-q^{-1})^2FE .
\ee
The algebra $U_{q}(\mathfrak{sl}(2,\CC))$ admits a real form
defined by the  involution
\be\label{real}
K^*=K,\,\ \ E^*=-E,\,\ \ \ F^*=-F,
\ee
which is compatible with the commutation relations (\ref{uqsl2}) only
if $|q|=1$, i.e. $\tau\in\RR$. The corresponding real algebra is
called $U_{q}(\mathfrak{sl}(2,\RR))$. (We shall see later that when
$U_{q}(\mathfrak{sl}(2))$ is replaced with its modular double,
there is a possibility to choose the real structure in
a different way.)

It is sometimes useful to consider the corresponding
`infinitesimal' algebra $U_\tau(\mathfrak{sl}(2,\RR))$, $\tau\in\RR$
with generators $E,\,F,\,H$ and relations
\be\label{utausl2}
[H,\,E]=2E,\ \ \ \ \  [H,\,F]=-2F, \\

[E,\,F]=\frac{q^{H}-q^{-H}}{q-q^{-1}}.
\ee
Evidently, there is an involution
\be\label{real2}
H^\ast=-H,\,\ \ E^*=-E,\,\ \ \ F^*=-F.
\ee

Let us sketch the representation theory of
$U_{q}(\mathfrak{sl}(2,\RR))$ in the way which stresses the role
of the modular duality concept (cf. \cite{PT}).
The representations of the principal series of
$U_{q}(\mathfrak{sl}(2,\RR))$ admit an explicit realization
by means of finite difference operators on the real line; the
commutation relations of the basic operators which are the
building blocks for these  representations are the ordinary Weyl
relations. To put it in a different way, the principal series
representations of $U_{q}(\mathfrak{sl}(2,\RR))$ factor through
a noncommutative torus.
\begin{de}
The noncommutative torus ${\AA}_{q}$ is the associative algebra
generated by $u,v$ subject to the relation $uv=q^2vu$.
\end{de}
We shall adjoin to  $\AA_{q}$ the inverse elements
$u^{-1},\,v^{-1},\,  $ (in other words, we replace $\AA_{q}$ with
its field of fractions, which we denote by the same letter).
\begin{prop}\label{GK}
For any $z\in{\CC}$ the mapping
$U_{q}(\mathfrak{sl}(2,\RR))\rightarrow\AA_{q}$ defined by
\be\label{GK1}
K \mapsto zu^{-1},\ \ \
E\mapsto\frac{v^{-1}}{q-q^{-1}}(1-u^{-1}),
\ \ \ F\mapsto\frac{qv}{q-q^{-1}}(z-z^{-1}u)
\ee
is a homomorphism of algebras.
\end{prop}
Note that the Casimir $C_2$ is mapped by the homomorphism (\ref{GK1})
to $qz+q^{-1}z^{-1} $.

It is sometimes technically convenient to extend the algebra
$U_q(\mathfrak{sl}(2))$ by adjoining to it `virtual Casimir elements'.
The following assertion is well-known.
\begin{prop}
The center of $U_q(\mathfrak{sl}(2,\RR))$ is isomorphic to the
polynomial algebra
$Z=\CC[qz+q^{-1}z^{-1}]\subset\CC[z, z^{-1}]=\hat{Z}$;
$U_q(\mathfrak{sl}(2,\RR))$ is a free $Z$-module.
\end{prop}
Set
$$
\hat{U}_q(\mathfrak{sl}(2,\RR))=
U_q(\mathfrak{sl}(2,\RR))\otimes_{Z}\hat{Z}.
$$
The mapping (\ref{GK1}) canonically extends to
$\hat{U}_q(\mathfrak{sl}(2))$. Informally, we may think of
$\hat{U}_{q}(\mathfrak{sl}(2))$ as of a bundle of noncommutative tori
parameterized by the spectrum of the central element $z\in\hat{Z}$.

Proposition \ref{GK} is a simple instance of the `free field
representations' for quantum groups; it may also
be compared with the well-known Gelfand-Kirillov \cite{GK1}
theorem which asserts that the field of fractions of the universal
enveloping algebra is isomorphic to the standard noncommutative
division algebra (central extension of the field of fractions of the
Weyl algebra generated by several pairs of `canonical variables'
$p_i,\,q_i$).

As a motivation for the study of the modular duality for
$U_{q}(\mathfrak{sl}(2,\RR))$ let us recall the following standard
construction from ergodic theory \cite {R}, \cite{Fad2}.
Let $q=\exp\pi i\omega_{1}/\omega _{2}$, where
$\omega_{1},\omega _{2}\in {\RR}$; we shall assume that
$\tau=\omega _{1}/\omega_{2}$ is irrational. Put
$\widetilde{q}=\exp(\pi i\omega_{2}/\omega _{1})$ and
let $\AA_{\widetilde{q}}$ be the dual torus with generators
$\widetilde{u},\,\widetilde{v}$, and relations
$\widetilde{u\!}\,\widetilde{v\!}\,=
\widetilde{q}^{\;2}\widetilde{v\!}\,\widetilde{u\!}$.
Let us define unitary operators
$T_{\omega _{1}},T_{\omega_{2}},S_{-i\omega_{1}},S_{-i\omega_{2}}$
in $L_{2}({\RR})$ by
\be
\begin{array}{ll}
T_{\omega _{1}}\varphi(t)=\varphi (t+\omega _{1}),\ \ \ \ \
& T_{\omega_{2}}\varphi(t)=\varphi (t+\omega _{2}),\\
S_{-i\omega _{1}}\varphi (t)=e^{\frac{2\pi it}{\omega _{1}}}\varphi(t),
& S_{-i\omega _{2}}\varphi(t)=e^{\frac{2\pi it}{\omega_{2}}}\varphi(t).
\end{array}
\ee
Define the dual representations of $\AA_{q}$ and
$\AA_{\wt q}$ in ${\mathcal H}=L_{2}({\RR})$ by
\be\label{tore}
\begin{array}{ll}
\rho:\ \  u\mapsto T_{\omega_{1}}, \ \ \ \ &
v\mapsto S_{-i\omega _{2}},\\
\widetilde{\rho}:\ \ \widetilde{u}\mapsto T_{\omega _{2}},
 & \widetilde{v} \mapsto S_{-i\omega _{1}}.
\end{array}
\ee
It is easy to see that $\AA_{q},$ and $\AA_{\wt q}$ are
the centralizers of each other in the algebra
${\mathcal B}({\mathcal H})$ of all bounded operators in
${\mathcal H}.$ The space ${\mathcal H}= L_2({\RR})$, which has the
structure of a left $\AA_{q}$-module and of a right
${\AA}_{\wt q}$-module is called the {\em imprimitivity
$({\AA}_q,$ $\AA_{\wt q})$-bimodule.}
The images of $\AA_{q},$ $\AA_{\wt q}$ in ${\mathcal B}({\mathcal H})$
are factors of type ${\rm II}_{1}$. Clearly, the  representations of
$\AA_q$ and $\AA_{\wt q}$ are reducible (in fact, both
$\AA_q$ and $\AA_{\wt q}$ contain plenty of
idempotent elements which are represented by projection operators
in ${\mathcal H};$ the image of a projection operator
$\wt P\in\wt\rho\left(\AA_{\wt q}\right)$
is an invariant subspace for $\AA_q$; the subspaces of
${\mathcal H}$ which arise in this way are the celebrated fractional
dimensional spaces of von Neumann).
On the other hand, the second commutant of
$\AA_q\otimes\AA_{\wt q}$ coincides with
${\mathcal B}({\mathcal H})$ and hence (\ref{tore}) is an irreducible
representation of $\AA_{q}\otimes \AA_{\wt q}$
(as a matter of fact, up to unitary equivalence, this algebra has a
unique irreducible representation).

The relation between the two noncommutative tori described above
is called by Rieffel the {\em strong Morita equivalence;}  In a
more general way, Rieffel showed \cite{R} that two tori $\AA_q$ and
$\AA_{\wt q}$, $q=e^{\pi i\tau},\wt q= e^{\pi i\wt \tau}$ are
strong Morita equivalent if and only if
$\wt\tau=\frac{a\tau+b}{c\tau+d},$ where
$$
\begin{pmatrix}
a & b\\
c & d
\end{pmatrix}\in GL(2,\ZZ),
$$
which explains the term `modular duality'.

\begin{de}
The modular dual of $U_{q}(\mathfrak{sl}(2,\RR))$ is the Hopf algebra
$U_{\wt q}(\mathfrak{sl}(2,\RR))$ with $\wt q= e^{\pi i/\tau}$;
we set also
$$
\hat U_{\wt q}\,(\mathfrak{sl}(2))=
U_{\wt q}\,(\mathfrak{sl}(2))\otimes_{\wt Z}\hat Z,
$$
$$
\wt Z=\CC[\wt qz+{\wt q}^{\;-1}z^{-1}], \ \ \ \
\hat Z= \CC[z,z^{-1}].
$$
\end{de}
The obvious motivation for this definition is the existence of the
`dual free field representation'
$U_{\wt q}(\mathfrak{sl}(2,\RR))\rightarrow\AA_{\wt q}$.
\begin{rem}
The modular transformation usually considered in the theory of
theta functions is $\tau\mapsto -\frac{1}{\tau} $; this
transformation preserves the upper half-plane ${\rm Im}\,\tau >0$
and the unit circle $|q|<1$. While the flip $\widetilde{q}\mapsto
\widetilde{q}^{\;-1}$ amounts to the simple exchange of the
generators of the quantum torus (and hence, in particular, the
quantum algebras $U_{\widetilde q}(\mathfrak{sl}(2,\RR))$ and
$U_{{\widetilde q}^{-1}}(\mathfrak{sl}(2,\RR))$ are
isomorphic), our choice of the sign of the modular transform
appears to be most natural for the study of $q$-deformed Toda chain.
\end{rem}

The fundamental difference which arises in the representation theory
of $U_{q}(\mathfrak{sl}(2,\RR))$ is that its unitary representations
are constructed from \emph{non-unitary} representations of the
quantum torus: we need to make a kind of 'Wick rotation' and hence
$u,v\in\AA_{q}$ are represented by \emph{unbounded} operators
\cite {S}.
More precisely, let us consider the following operators in
${\mathcal H}$:
\be\label{op}
\begin{array}{ll}
T_{i\omega_{1}}\varphi(t)=\varphi(t+i\omega_{1}),\ \ \ \ \ \ &
T_{i\omega_{2}}\varphi(t)=\varphi (t+i\omega _{2}),\\
S_{\omega_{1}}\varphi(t)=e^{\frac{2\pi t}{\omega_{1}}}\varphi(t),
&
S_{\omega_{2}}\varphi(t)=e^{\frac{2\pi t}{\omega_{2}}}\varphi(t).
\end{array}
\ee
The dual representations of $\AA_{q}, \ \AA_{\widetilde{q}}$
are now given by
\be
\begin{array}{ll}
\rho_{W}: \ \ \ u \mapsto T_{i\omega _{1}},\ \ \ \
& v\mapsto S_{\omega _{2}},\\
\widetilde{\rho}_{W}: \ \ \ \widetilde{u}\mapsto T_{i\omega _{2}},
& \widetilde{v}\mapsto S_{\omega _{1}}.
\end{array}
\ee
Operators (\ref{op}) are essentially  self-adjoint on the common
domain ${\mathcal P}$ which consists of entire functions $\psi $
such that
$$
\int\limits_{\rs} e^{sx}\left| \psi \left( x+iy\right)
\right|^2 dx <\infty \ \ \ \ \ \mbox{for all}\,y\in\RR,\ \ s\in\RR.
$$
\begin{rem}
Unlike the unitary case, the definition of the centralizer of an
unbounded operator must take care of the domains of operators;
thus $AB=BA$ implies that $B({\rm Dom}_A)\subset{\rm Dom}_A$; this may
be not true even if $B$ is bounded. As a result, although the four
operators (\ref{op}) commute with each other, the same is not true,
e.~g., for their spectral projection operators
\footnote{
Spectral projection operators $E(\Delta)$ for multiplication operators
are multiplication operators by the characteristic function of the
interval $\Delta$; this function has compact support and hence the
spectral projector does not preserve the domain which consists of
analytic functions.
};
thus, contrary to the ergodic case, the centralizer of
$\rho _{W}\left(\AA_{q}\right)$ does not contain projection
operators, and hence the representations
$\rho_{W},\widetilde{\rho}_{W}$ are geometrically irreducible.
It is much more important for us, however, that they are
\emph{not irreducible} in the operator sense, as each
of them still admits a huge algebra of intertwiners.
\end{rem}

\begin{prop}\label{centr}
Operators which commute with all four operators (\ref{op}) are
scalars.
\end{prop}

\begin{cor}
Representation of $\AA_{q}\otimes\AA_{\widetilde{q}}$
is strongly irreducible (i.e. it does not admit any nontrivial
intertwiners$).$
\end{cor}
Let us now describe explicitly the {\it particular}
principal series representation of
$\hat{U}_{q}(\mathfrak{sl}(2,\RR))$
which is extensively used in the paper (the realization we use
is slightly different from those described in \cite{PT}).
Namely, the representation ${\pi }_{\lambda }$ of
$\hat{U}_{q}(\mathfrak{sl}(2,\RR))$
($q=e^{\pi i\omega_1/\omega_2}\,,\omega_1,\omega_2\in\RR_+$),
which depends on a parameter $\lambda\in{\CC},$ is given by
\be\label{omega1}
\pi_\lambda:\ \ \ \ \ \ \
\begin{array}{l}
K\mapsto e^{\frac{\pi i\lambda}{\omega_2}}
T_{i\omega_1}^{-1}, \\
E\mapsto\frac{S_{\omega_2}^{-1}}{q-q^{-1}}
\left(1-T_{i\omega_1}^{-1}\right),\\
F\mapsto\frac{qS_{\omega_2}}{q-q^{-1}}
\left(e^{\frac{\pi i\lambda}{\omega_2}}
-e^{-\frac{\pi i\lambda}{\omega_2}}T_{i\omega_{1}}\right),\\
C_2\mapsto qe^{\frac{\pi i\lambda}{\omega_2}}+
q^{-1}e^{-\frac{\pi i\lambda}{\omega_2}}, \\
z\mapsto e^{\frac{\pi i\lambda}{\omega_2}}.
\end{array}
\ee
By duality, we define the representation $\widetilde\pi_\lambda$
of the modular dual algebra
$\hat{U}_{\widetilde{q}}(\mathfrak{sl}(2,\RR))$
with $\wt q=e^{\pi i\omega_2/\omega_1}$ by
\be\label{omega2}
\wt\pi_\lambda:\ \ \ \ \ \ \
\begin{array}{l}
\wt K\mapsto e^{\frac{\pi i\lambda }{\omega _1}}
T_{i\omega _2}^{-1}, \\
\wt E\mapsto\frac{S_{\omega_1}^{-1}}{\wt q-\wt q^{\;-1}}
\left(1-T_{i\omega_2}^{-1}\right),\\
\wt F\mapsto\frac{\wt q\,S_{\omega_1}}{\wt q-\wt q^{\;-1}}
\left(e^{\frac{\pi i\lambda}{\omega_1}}
-e^{-\frac{\pi i\lambda }{\omega_1}}
T_{i\omega_2}\right),\\
\wt C_2\mapsto \wt q\,e^{\frac{\pi i\lambda}{\omega_1}}+
\wt q^{\;-1}e^{-\frac{\pi i\lambda}{\omega_1}}, \\
\wt z\mapsto e^{\frac{\pi i\lambda}{\omega_1}}.
\end{array}
\ee
The representations  $\pi_\lambda $, $\widetilde{\pi }_{\lambda }$
are defined on a larger space ${\cal P}_{\lambda}\supset {\cal P}$
which depends on $\lambda$.
\begin{de}
${\cal P}_{\lambda}$ is the set of entire functions such that
\renewcommand{\theenumi}{(\roman{enumi})}
\begin{enumerate}
\item
For $t\rightarrow+\infty$ a function $\psi\in {\cal P}_{\lambda}$
admits an asymptotic expansion
\be\label{test}
\psi (t+is) \sim_{t\rightarrow + \infty} e^{\frac{2\pi\lambda t}
{\omega_1\omega_2}}\sum_{n_{1},n_{2}\geq 0}C_{n_{1},n_{2}}
e^{\frac{-2\pi t(n_{1}\omega_1+n_{2}\omega_2)}{\omega_1\omega_2}}
\ee
uniformly in each bounded strip.
\item
For $t\rightarrow - \infty$ it admits an asymptotic expansion
\be
\psi (t+is) \sim_{t\rightarrow - \infty} C\Big(1+\sum_{n_{1},n_{2} >0}
C_{n_{1},n_{2}}e^{\frac{2\pi t(n_{1}\omega_1+ n_{2}\omega_2)}
{\omega_1\omega_2}}\Big)
\ee
uniformly in each bounded strip
\footnote{
The space ${\cal P}_\la$ essentially coincides with those considered
in \cite{PT}.
}
.
\end{enumerate}
\end{de}
The scalar product which is adapted to the discussion of the
unitarity conditions in our algebra, defined on ${\cal P}_{\lambda}$
only for ${\lambda}\in i\RR-\omega_1-\omega_2$, is given by
\be\label{sp1}
(\varphi,\psi) =\int\limits_{\rs}
e^{2\pi t\left(\frac{1}{\omega _{1}}+\frac{1}{\omega_{2}}\right)}
\ov{\varphi(t)\!\!}\;\psi(t)dt
\ee
\begin{prop}\label{P lambda} The following statements hold:
\begin{itemize}
\item[(i)] Operators $\pi_\lambda(X),\,
 X\in U_q(\mathfrak{sl}(2,\RR))$
and $\wt\pi_\lambda(Y),\,Y\in U_{\wt q}\,(\mathfrak{sl}(2,\RR))$
leave ${\cal P}_\lambda\subset L_2(\RR)$ invariant and
commute with each other on this domain; any operator which
commutes with both algebras is scalar.
\item[(ii)]
If $\lambda\in i\RR-\omega_1-\omega_2$, all
operators $\pi_\lambda (X),\ X\in U_q(\mathfrak{sl}(2,\RR))$
obey the involution generated by (\ref{real})
with respect to the scalar product (\ref{sp1}). Similar statement
holds for all operators
$\wt\pi_\lambda(Y),\ Y\in U_{\wt q}(\mathfrak{sl}(2,\RR))$.
\item[(iii)]
The central characters $z,\wt z$ of $\pi_\lambda, \ \wt \pi_\lambda$
are related by $z=\wt z^{\;\tau}, \ \tau=\omega_1/\omega_2 $.
\end{itemize}
\end{prop}
As in Proposition \ref{centr}, commutativity condition implies
that the operator preserves the domains of our unbounded
operators.
\begin{cor}\label{modular}
The principal series representation $\pi_\lambda$ of
$\hat{U}_q(\mathfrak{sl}(2,\RR))$ canonically extends to a
representation of
$\hat{U}_{{q}}(\mathfrak{sl}(2,\RR))\otimes\hat U_{\wt q}
(\mathfrak{sl}(2,\RR))$ which is defined on the same domain
${\cal P}_\lambda$; this representation is unitary if and only if
$\lambda\in i\RR-\omega_1-\omega_2$.
\end{cor}
\begin{rem}
Since representations of the dual algebras $ U_q(\mathfrak{sl}(2))$
and $U_{\wt q}\,(\mathfrak{sl}(2))$ are constructed from the dual
representations of the quantum tori $\AA_q,\ \AA_{\wt q}$ for any
values of the central characters
$z=e^{\pi i\lambda/\omega_2}, \wt z=e^{\pi i\mu/\omega_1}$, one might
conclude that the principal series representations
$\pi_\lambda,\,\wt\pi_\mu$ centralize each other for any pair of
indices $\lambda, \ \mu$; it is  the condition on the common domain
which imposes the selection rule.
\end{rem}
Let us now introduce the following key definition.
\begin{de}
The modular double of $ \hat{U}_q(\mathfrak{sl}(2))$ is the
Hopf algebra
$$
{\mathcal D}_{mod}=\hat{U}_q(\mathfrak{sl}(2))\otimes
\hat{U}_{\wt q}\,(\mathfrak{sl}(2)).
$$
\end{de}
The bialgebra structure of the modular double, i.e., its product
and coproduct, is standard. The point is that this algebra admits
an unexpected class of representations which are \emph{not}
tensor products of representations of the factors, but rather are
related to a kind of ``type II'' operator algebras (the quotation
marks reflect the fact that due to analyticity constraints our
algebras are ``thinner'' than the genuine type II factors; in
particular, they do not contain projection operators).  The
modular double should be regarded as an analytic rather than
algebraic object which for the first time brings into play the
nontrivial analytic properties of noncompact semisimple quantum
groups.

In what follows we shall be interested only in the {\em  principal
series representations} of ${\mathcal D}_{mod}$ defined above;
with respect to this subclass of representations
${\mathcal D}_{mod}$ behaves itself as a rank one algebra.
Note that the kernel of these representations contains the
two-sided ideal ${\mathcal J}\subset\hat{U}_{{q}}(\mathfrak{sl}(2))
\otimes\hat{U}_{\wt q}(\mathfrak{sl}(2))$ generated by the relations
$$
z=\wt z^{\;\tau}, \ \ \  K=\wt K^{\;\tau}.
$$

The use of the modular double and its representations, instead
of those of its factors, appears to be very natural in many ways.
We shall see below that the definition of the Whittaker vectors
becomes unambiguous only if we require that they are the
eigenvectors of \emph{both} nilpotent generators
$\pi_\lambda(E),\;\wt\pi_{\lambda}(\wt E)$. One more reason to enjoy
the presence of a double set of generators is the integrability
problem for the $q$-deformed (relativistic) Toda model discussed below.
The $q$-Toda Hamiltonian,
which is derived from the Casimir element of $U_q(\mathfrak{sl}(2))$
is a difference operator which involves only translations
$T_{i\omega_1}$; due to the presence of quasiconstants (i.e.,
functions with period $i\omega_1$), its spectrum becomes multiple
with infinite multiplicity; the multiplicity problem is resolved
when we take into account the dual Casimir element which involves
dual translations $T_{i\omega_2}$.

The real form of ${\mathcal D}_{mod}$ used above is inherited from
the real forms of $\hat U_q(\mathfrak{sl}(2)), \
\hat U_{\wt q}(\mathfrak{sl}(2))$. Interestingly, there exists a
kind of complexification of the picture described above.
As pointed out by Faddeev
\cite{Fad2}, for a special choice of the {\it complex} periods
$\omega_1, \ \omega_2$ there exists another real form of
${\mathcal D}_{mod}$ which does not reduce to real forms of its
factors. Namely,
\begin{prop}
\begin{itemize}
\item[(i)]
Let us assume that $\omega_1 =\ov{\omega}_2$, or,
equivalently, that $|\tau|=1$. Then the mapping
\be\label{freal}
E\mapsto -\wt E, \; F\mapsto -\wt F, \; K\mapsto \wt K, \;
z\mapsto{\wt q}^{\;2}\wt z
\ee
extends to a $\CC$-antilinear involution of ${\mathcal D}_{mod}$.
\item[(ii)]
Let $\rho$ be a unitary representation of ${\mathcal D}_{mod}$
with respect to the real form (\ref{freal}); then all operators
$\rho(X), X\in \hat{U}_q(\mathfrak{sl}(2))\subset{\mathcal D}_{mod},
\,\rho(Y), Y\in \hat{U}_{\wt q}(\mathfrak{sl}(2))
\subset{\mathcal D}_{mod}$,
are normal.
\item[(iii)]
Let $\lambda\in i\RR-\omega_1-\omega_2$;
then the principal series representation $\pi_\lambda$ extends to
a unitary representation of ${\mathcal D}_{mod}$ with respect to
the real form (\ref{freal}).
\end{itemize}
\end{prop}
Physical self-adjoint Hamiltonians associated with the real form
(\ref{freal}) can be derived from the real and imaginary parts
of the Casimir operators. Analytically, the Faddeev's real form is
particularly attractive, since in that case the lattice generated
by $\omega _{1},\omega_{2}$ is non-degenerate.

The use of the modular double is very well suited for the
treatment of interpolation problems. Recall that we are dealing
with the `rational form' of the quantum algebra
$U_{q}(\mathfrak{sl}(2))$ which is defined in terms of the generator
$K=q^H$. This choice is at the core of the modular duality: it will
be completely destroyed if we replace $U_{q}(\mathfrak{sl}(2))$
with the 'infinitesimal' algebra $U_{\tau}(\mathfrak{sl}(2))$
generated by  $E,F,H$, the commutativity of two dual sets of
generators will be destroyed. On the other hand, when it comes up
to compute special functions associated with representations of
$U_{q}(\mathfrak{sl}(2))$, i.e., some specific matrix coefficients
of its irreducible representations, e.g., spherical functions or
Whittaker functions, and to construct the corresponding spectral
theory, it is important to define these functions on the entire real
line or on its compexification. By contrast, the use of the rational
form $U_{q}(\mathfrak{sl}(2))$ implies that these functions are
defined \emph{a priori } only on a discrete set $\{K^n,\,n\in\ZZ\}$.
Let assume that $\omega_1, \ \omega_2$ are real and
$\tau=\omega_1/\omega_2$ is irrational.
\begin{prop}
For any $\alpha\in\RR$ operators $\pi_\lambda(e^{\alpha H})$ are
approximated by linear combinations of
$\pi_\lambda(K^n\cdot\wt K^{\;m}), \ n,m\in\ZZ$.
\end{prop}
Indeed,$\pi_\lambda(e^{\alpha H})$ is a translation operator,
$\pi_\lambda(e^{\alpha H})\varphi(t)=\varphi(t-i\alpha)$; on the
other hand,
$\pi_\lambda(K^n\cdot\wt K^{\;m})\varphi(t)=
e^{\pi i\lambda n/\omega_2+\pi i\lambda m/\omega_1}
\varphi(t-in\omega_1-im\omega_2)$. The set
$\{in\omega_1+im\omega_2\ ;\ n,m\in\ZZ\}$ is dense in $i\RR$.
\begin{rem}
There exists a whole family of principal series
representations similar to those described above. It is easy to
find a realization of the algebras $U_q(\mathfrak{sl}(2))$
and $U_{\wt q}(\mathfrak{sl}(2))$ labeled by integer indices
$k_1$ and $k_2$, respectively, such that all representation
operators, which act on an appropriate space ${\cal P}^{k_1k_2}_\la$,
satisfy the unitarity condition with respect to the scalar product
with the measure
$\exp\{\frac{2\pi(k_1\omega_1+k_2\omega_2)t}{\omega_1\omega_2}\}$.
In this case one obtains unitary representation if and only if
$\la\in i\RR-k_1\omega_1-k_2\omega_2$. For simplicity, in the present
paper we restrict ourself to the case $k_1=k_2=1$, although
the more general case can be treated quite similarly.
\end{rem}

\section{Whittaker vectors}
Let $\mathfrak g$ be a semisimple Lie algebra, $\mathfrak n$ its
maximal nilpotent subalgebra generated by positive root vectors. A
character $ \chi :{\mathfrak n}\rightarrow\CC$ is uniquely
fixed by its values on root vectors associated with simple roots;
it is called nondegenerate if $ \chi (e_\alpha)\neq 0$ for all
simple roots $ \alpha$. A Whittaker vector in a  $\mathfrak
g$-module $V$ is a vector $w\in V$ such that
\begin{align}
Xw&= \chi(X)w
\end{align}
for all $X\in{\mathfrak n}$ The extension of this definition to
$q$-deformed algebras is nontrivial: it is easy to see that for
${\rm rank} \ {\mathfrak g}\geq 2$ the algebra $U_q({\mathfrak n})$
generated by the Chevalley generators associated with positive simple
roots does not admit nondegenerate characters (the obstruction is
associated with the $q$-deformed Serre relations). In \cite{sev}
Sevostyanov found the way around this difficulty: one has to rescale
the generators of the nilpotent subalgebra multiplying them by
appropriate group-like elements from the Cartan subalgebra.
Although the Serre relations are vacuous in the $\mathfrak{sl}_2$
case, the same trick proves worthy in that case as well; it
provides an extra freedom which serves to construct various
versions of the $q$-deformed Toda Hamiltonians.

Whittaker vectors associated with the unitary principal series
representations of $SL(2,\RR)$ do not lie in the Hilbert
space, because the spectrum of $E$, $F$ is continuous; as a
result, the Whittaker functions which are defined as formal matrix
coefficients of the principal series representations between a
pair of Whittaker vectors are expressed by a divergent integral
which requires regularization. The situation in the $q$-deformed
case is completely similar. As already mentioned, the natural
definition of  Whittaker vectors in the $q$-deformed setting
requires the use of the modular double. The two commuting
generators $E, \ \widetilde{E} \in {\mathcal D}_{mod}$ give rise
to two compatible difference equations which have a unique common
solution with nice analytic properties; this solution does not
belong to $ L_2(\RR)$, because it does not decrease rapidly
enough.

With these remarks in mind, we may now proceed to the formal
definition. Let $(\pi_\lambda,\wt\pi_\lambda)$,
$\lambda\in i\RR-\omega_1-\omega_2$
be the unitary representation of ${\mathcal D}_{mod}$
and $\al\in\RR$ be an arbitrary parameter. The
$E$-Whittaker vector $ \Phi^{(\al)}_{\lambda}$ is defined by
\be\label{gv1}
\pi_\lambda(E)\Phi^{(\al)}_{\lambda}=
\frac{\zen^{\omega_1}}{q-q^{-1}}\,
e^{\pi i\al}\,\pi_\lambda (q^{\al H})\Phi^{(\al)}_{\lambda}, \\
\wt\pi_\lambda(\widetilde{E})\Phi^{(\al)}_{\lambda}=
\frac{\zen^{\omega_2}}{\wt q-\wt q^{\,-1}}\,e^{\pi i\al}\,
\wt\pi_\lambda(\wt q^{\;\al\wt H})\Phi^{(\al)}_{\lambda};
\ee
here $\zen$ is a positive real number (the `coupling constant').
The extra parameter $\alpha$ matches the freedom in the choice of
the quantum Lax operator in the alternative formulation of the
$q$-deformed Toda theory based on Quantum Inverse Scattering Method.
In other words, particular choices of $\alpha$ correspond
to different Toda-like models.

In a similar way, the $F$-Whittaker vectors are defined by
\be\label{gv1b}
\pi_\lambda(F)\wh\Phi^{(\al)}_{\lambda}=-
\frac{\zen^{\omega_1}}{q-q^{-1}}\, e^{\pi i\al}\,
\pi_\lambda(q^{-\al H})\wh\Phi^{(\al)}_{\lambda},\\
\wt\pi_\lambda(\wt F)\wh\Phi^{(\al)}_{\lambda}=-
\frac{\zen^{\omega_2}}{\wt q-\wt q^{\,-1}}\,e^{\pi i\al}\,
\wt\pi_\lambda(\wt q^{\;-\al\wt H})
\wh\Phi^{(\al)}_{\lambda}.
\ee

The definition of the Whittaker vectors is completely symmetric
with respect to the exchange of the two dual algebras
$U_q(\mathfrak{sl}(2)),\ U_{\wt q}\,(\mathfrak{sl}(2))$.
Note that the existence of a common
eigenvector of the commuting generators $\pi_\lambda(E), \
\wt\pi_\lambda(\wt E),$ or $\pi_\lambda(F), \
\wt\pi_\lambda(\wt F)$ is guaranteed due to our `selection
rule' for the central characters of $\hat{U}_q(\mathfrak{sl}(2)), \
\hat{U}_{\wt q}\,(\mathfrak{sl}(2))$.

\subsection{Whittaker vectors: Explicit solutions}
We shall start with the explicit formulae for the
simplest Whittaker vectors corresponding to a particular choice
of $\al$. Using the representations (\ref{omega1}),
(\ref{omega2}), we get the following system of difference
equations for the vectors $\Phi^{(\al)}_\la$ with $\alpha =0,1$:
\begin{subequations}\label{es1}
\be\label{es1a}
\frac{\Phi^{(0)}_\la(t-i\omega_1)}
{\Phi^{(0)}_\la(t)}=1-\zen^{\omega_1}e^{\frac{2\pi t}{\omega_2}},
\ee
\be\label{es1b}
\frac{\Phi^{(0)}_\la(t-i\omega_2)}
{\Phi^{(0)}_\la(t)}=1-\zen^{\omega_2}e^{\frac{2\pi t}{\omega_1}},
\ee
\be\label{es2a}
\frac{\Phi^{(1)}_\la(t-i\omega_1)}{\Phi^{(1)}_\la(t)}=
\frac{1}{1-\zen^{\omega_1}
e^{\frac{2\pi t}{\omega_2}+\frac{\pi i\la}{\omega_2}}}\;,
\ee
\be\label{es2b}
\frac{\Phi^{(1)}_\la(t-i\omega_2)}{\Phi^{(1)}_\la(t)}=
\frac{1}{1-\zen^{\omega_2}
e^{\frac{2\pi t}{\omega_1}+\frac{\pi i\la}{\omega_1}}}\;.
\ee
\end{subequations}
In a similar way, the Whittaker vectors
$\wh\Phi^{(\al)}_\la$ with $\al=0, 1$
satisfy the difference equations
\begin{subequations}
\be\label{es1c}
\frac{\wh\Phi^{(0)}_\la(t+i\omega_1)}
{\wh\Phi^{(0)}_\la(t)}=e^{\frac{2\pi i\la}{\omega_2}}
\Big\{1+q^{-1}\zen^{\omega_1}e^{-\frac{2\pi t}{\omega_2}-
\frac{\pi i\la}{\omega_2}}\Big\},
\ee
\be\label{es2c}
\frac{\wh\Phi^{(0)}_\la(t+i\omega_2)}
{\wh\Phi^{(0)}_\la(t)}=e^{\frac{2\pi i\la}{\omega_1}}
\Big\{1+\widetilde{q}^{\;-1}
\zen^{\omega_2}e^{-\frac{2\pi t}{\omega_1}-
\frac{\pi i\la}{\omega_1}}\Big\},
\ee
\be\label{es1d}
\frac{\wh\Phi^{(1)}_\la(t+i\omega_1)}{\wh\Phi^{(1)}_\la(t)}=
\frac{e^{\frac{2\pi i\la}{\omega_2}}}
{1+q^{-1}\zen^{\omega_1}e^{-\frac{2\pi t}{\omega_2}}}\;,
\ee
\be\label{es2d}
\frac{\wh\Phi^{(1)}_\la(t+i\omega_2)}{\wh\Phi^{(1)}_\la(t)}=
\frac{e^{\frac{2\pi i\la}{\omega_1}}}
{1+{\wt q}^{\,-1}\zen^{\omega_2}
e^{-\frac{2\pi t}{\omega_1}}}\;.
\ee
\end{subequations}
Let ${\cal S}(y)$ be the function defined in terms of the double
sine $S_2(y)$ according to (\ref{cs}).\footnote{ In the main text
we shall write ${\cal S}(y)$ instead of ${\cal S}(y|\bo)$ for
brevity. We omit such dependence for any other function of such
type. }

\begin{prop}
The Whittaker vectors satisfying the equations (\ref{es1a} --
\ref{es2d}) are given by the following formulae:
\begin{subequations}\label{es3}
\be\label{es3b}
\Phi^{(0)}_\la(t)={\cal S}\Big(\!-it+\omega_1\!+\!\omega_2-
{\textstyle\frac{i\omega_1\omega_2}{2\pi}}\log\zen\Big),
\ee
\be\label{es3a}
\Phi^{(1)}_\la(t)={\cal S}^{-1}\Big(\!-it+
\omega_1\!+\!\omega_2\!+\!\frac{\la}{2}-
{\textstyle\frac{i\omega_1\omega_2}{2\pi}}\log\zen\Big),
\ee
\be\label{es3c}
\wh\Phi^{(0)}_\la(t)={\cal S}\Big(it+
{\textstyle\frac{1}{2}}(\omega_1\!+\!\omega_2)-\frac{\la}{2}
-{\textstyle\frac{i\omega_1\omega_2}{2\pi}}\log\zen\Big)
e^{\frac{2\pi\la t}{\omega_1\omega_2}},
\ee
\be\label{es3d}
\wh\Phi^{(1)}_\la(t)=
{\cal S}^{-1}\Big(it\!+\!{\textstyle\frac{1}{2}}
(\omega_1\!+\!\omega_2)-
{\textstyle\frac{i\omega_1\omega_2}{2\pi}}\log\zen\Big)
e^{\frac{2\pi\la t}{\omega_1\omega_2}}.
\ee
\end{subequations}
\end{prop}

In a more general way, one can prove the following formulae for
the Whittaker vectors $ \Phi_\la^{(\alpha)}, \
\wh\Phi_\la^{(\alpha)}$ with arbitrary values of $\alpha$:
\begin{subequations}\label{gv3}
\be\label{gv3a}
\Phi^{(\al)}_\la(t)=\int\limits_{\Gamma_\al}c(\zeta)
e^{\frac{\pi i(2\al-1)\zeta^2}{2\omega_1\omega_2}+
\frac{2\pi i\zeta}{\omega_1\omega_2}
[t+\frac{i\al}{2}(\la+\omega_1+\omega_2)+
\frac{i}{4}(\omega_1+\omega_2)]}\,d\zeta,
\ee
\be\label{gv3b}
\wh\Phi^{(\al)}_\la(t)=e^{\frac{2\pi\la t}{\omega_1\omega_2}}
\int\limits_{\Gamma_\al}c(\zeta)
e^{\frac{\pi i(2\al-1)\zeta^2}{2\omega_1\omega_2}-
\frac{2\pi i\zeta}{\omega_1\omega_2}
[t+\frac{i(1-\al)}{2}(\la+\omega_1+\omega_2)-
\frac{i}{4}(\omega_1+\omega_2)]}\,d\zeta,
\ee
\end{subequations}
where
\be\label{ch1}
c(\zeta)\ \equiv \
\frac{\zen^{i\zeta}}{\sqrt{\omega_1\omega_2}}\,S^{-1}_2(-i\zeta)
\ee
and the contour  $\Gamma_\al$ is chosen in such a way that
it passes above the poles of the integrand and escapes to infinity
in the sector where the function
$e^{\frac{\pi i\al\zeta^2}{\omega_1\omega_2}}$
is decaying on the left and in the sector where
$e^{\frac{\pi i(\al-1)\zeta^2}{\omega_1\omega_2}}$ is decaying
on the right. For $\alpha \neq 0, 1$
$\Phi^{(\alpha)}_\lambda, \ \wh\Phi^{(\alpha)}_\lambda$ are entire
functions of the variable $t$; for "degenerated cases"
$\alpha =0, 1$ the integrals in
\eqref{gv3a}, \eqref{gv3b} may be evaluated explicitly using the
formulae \eqref{ft1} and reduce to \eqref{es3};
in these cases both vectors are meromorphic functions of $t$.

Let us note that the function $c(\zeta)$ may be regarded as
the {\em $q$-deformed Harish-Chandra function} (this term is
justified by its role in the asymptotic formulae for the Whittaker
functions, see below).

\subsection{Whittaker functions}
Now we would like to define the $q$-deformed Whittaker
functions as the matrix elements of Whittaker vectors. As it was
mentioned before, the standard integral (\ref{sp1}) is divergent in
this case. To
regularize the integral, one should deform the integration contour
in an appropriate way. Therefore, by the scalar product below we mean
the suitable regularization of (\ref{sp1}).

\begin{de}
Let $\bal=(\al_1,\al_2)\in\RR^2$.
The Whittaker functions $w^{(\bal)}_\la(x)$ corresponding to
representation $(\pi_\la,\,\wt\pi_\la)$ of the
algebra ${\mathcal D}_{mod}$ are the matrix elements
\be\label{wf6}
w^{(\bal)}_\la(x)=
e^{-\frac{\pi(\omega_1+\omega_2)x}{\omega_1\omega_2}}
\Big(\wh\Phi^{(\alpha_1)}_\la,
e^{-\frac{\pi x}{\omega_2}H}\Phi^{(\alpha_2)}_\la\Big)
\ee
\end{de}

\begin{prop}
The Whittaker functions (\ref{wf6}) satisfy the equations
\begin{subequations}\label{gt1}
\be\label{gt1a}
\hspace{-0.3cm}
w^{(\bal)}_\la(x\!-\!i\omega_1)+
w^{(\bal)}_\la(x\!+\!i\omega_1)+q^{\al_1-\al_2}\zen^{2\omega_1}
e^{\frac{2\pi x}{\omega_2}}
w^{(\bal)}_\la(x\!+\!i(\al_1\!-\!\al_2)\omega_1)=\\
-\,\Big(qe^{\frac{\pi i\la}{\omega_2}}+
q^{-1}e^{-\frac{\pi i\la}{\omega_2}}\Big)w^{(\bal)}_\la(x),
\ee
\be\label{gt1b}
\hspace{-0.3cm}
w^{(\bal)}_\la(x\!-\!i\omega_2)+
w^{(\bal)}_\la(x\!+\!i\omega_2)+
\wt q^{\;\al_1-\al_2}\zen^{2\omega_2}e^{\frac{2\pi x}{\omega_1}}
w^{(\bal)}_\la(x\!+\!i(\al_1\!-\!\al_2)\omega_2)=\\
-\,\Big(qe^{\frac{\pi i\la}{\omega_1}}+
q^{-1}e^{-\frac{\pi i\la}{\omega_1}}\Big)w^{(\bal)}_\la(x).
\ee
\end{subequations}
\end{prop}

Let us check (\ref{gt1a}) formally; we shall discuss the convergence
of the integral in (\ref{wf6}) a little later. Set
\be\label{wf1}
{\cal F}^{(\bal)}_\la(x)=\Big(\wh\Phi^{(\al_1)}_\la,
e^{-\frac{\pi x}{\omega_2}H}\Phi^{(\al_2)}_\la\Big).
\ee
The eigenvalue of the Casimir operator $\pi_\lambda(C_2)$ is
\be\label{rep3}
C_2=qe^{\frac{\pi i\la}{\omega_2}}+
q^{-1}e^{-\frac{\pi i\la}{\omega_2}}.
\ee
Therefore,
\be\label{wf2}
\Big(\wh\Phi^{(\al_1)}_\la,
e^{-\frac{\pi x}{\omega_2}H}C_2\Phi^{(\al_2)}_\la\Big) =
(qe^{\frac{\pi i\la}{\omega_2}}+
q^{-1}e^{-\frac{\pi i\la}{\omega_2}}){\cal F}^{(\bal)}_\la(x).
\ee
On the other hand,
\be
\hspace{-0.3cm}
\Big(\wh\Phi^{(\al_1)}_\la,
e^{-\frac{\pi x}{\omega_2}H}C_2\Phi^{(\al_2)}_\la\Big)= \\
\Big(\wh\Phi^{(\al_1)}_\la,e^{-\frac{\pi x}{\omega_2}H}
\Big\{q^{H+1}+q^{-H-1}+(q\!-\!q^{-1})^2FE\Big\}
\Phi^{(\al_2)}_\la\Big)= \\
\hspace{-0.75cm}
\Big(\wh\Phi^{(\al_1)}_\la,e^{-\frac{\pi x}{\omega_2}H}
(q^{H+1}\!+q^{-H-1})\Phi^{(\al_2)}_\la\Big)
-(q\!-\!q^{-1})^2e^{\frac{2\pi x}{\omega_2}}
\Big(F\wh\Phi^{(\al_1)}_\la,
e^{-\frac{\pi x}{\omega_2}H}E\Phi^{(\al_2)}_\la\Big).
\ee
Using the definition of the Whittaker vectors (\ref{gv1}),
(\ref{gv1b}), we obtain
\be\label{wf2'}
\Big(\wh\Phi^{(\al_1)}_\la,
e^{-\frac{\pi x}{\omega_2}H}C_2\Phi^{(\al_2)}_\la\Big)=
\Big(\wh\Phi^{(\al_1)}_{\lambda},
e^{-\frac{\pi x}{\omega_2}H}(q^{H+1}+q^{-H-1})
\Phi^{(\al_2)}_\la\Big)-\\
e^{\pi i(\al_2-\al_1)}\zen^{2\omega_1}e^{\frac{2\pi x}{\omega_2}}
\Big(\wh\Phi^{(\al_1)}_\la,e^{-\frac{\pi x}{\omega_2}H}
q^{(\al_2-\al_1)H}\Phi^{(\al_2)}_\la\Big)=\\
\Big\{qe^{-i\omega_1\d_{x}}+q^{-1}e^{i\omega_1\d_{x}}-
e^{\pi i(\al_2-\al_1)}\zen^{2\omega_1}e^{\frac{2\pi x}{\omega_2}}
e^{i(\al_1-\al_2)\omega_1\d_x}\Big\}{\cal F}^{(\bal)}_\la(x)
\ee
From (\ref{wf2}) and (\ref{wf2'}) it follows that the matrix
coefficient ${\cal F}^{(\bal)}_\la$ satisfies to equation
\be\label{wf3}
\Big\{qe^{-i\omega_1\d_{x}}+q^{-1}e^{i\omega_1\d_{x}}-
e^{\pi i(\al_2-\al_1)}\zen^{2\omega_1}e^{\frac{2\pi x}{\omega_2}}
e^{i(\al_1-\al_2)\omega_1\d_x}\Big\}{\cal F}^{(\bal)}_\la(x)=
\\ \Big(qe^{\frac{\pi i\la}{\omega_2}}+
q^{-1}e^{-\frac{\pi i\la}{\omega_2}}\Big){\cal F}^{(\bal)}_\la(x)
\ee
Hence, the function
\be\label{wf4}
w^{(\bal)}_\la(x )=
e^{-\frac{\pi(\omega_1+\omega_2)x} {\omega_1\omega_2}}
{\cal F}^{(\bal)}_\la(x)
\ee
satisfies (\ref{gt1a}).\square
\begin{cor}
Let the unitary weight be $\la=-i\gamma\!-\!\omega_1\!-\!\omega_2$.
The Whittaker functions
$w^{(\bal)}_{-i\gamma-\omega_1-\omega_2}\equiv w^{(\bal)}_\gamma$ are
eigenfunctions of the Hamilton operators
\be
{\mathcal H}^{(\al_1-\al_2)}=e^{i\omega_1\partial_x}\ +\
e^{-i\omega_1\partial_x}+
q^{\al_1-\al_2}\zen^{2\omega_1}e^{\frac{2\pi x}{\omega_2}}
e^{i(\al_1-\al_2)\omega_1\partial_x}, \\
\widetilde{{\mathcal H}}^{(\al_1-\al_2)}=e^{i\omega_2\partial_x}\ +\
e^{-i\omega_2\partial_x}+\widetilde{q}^{\;\al_1-\al_2}\zen^{2\omega_2}
e^{\frac{2\pi x}{\omega_1}}e^{i(\al_1-\al_2)\omega_2\partial_x}
\ee
with eigenvalues $
\varepsilon_\gamma = e^{\frac{\pi\gamma}{\omega_2}} +
e^{-\frac{\pi\gamma}{\omega_2}}$ and
$ \wt\varepsilon_\gamma =
e^{\frac{\pi\gamma}{\omega_1}} + e^{-\frac{\pi\gamma}{\omega_1}}$,
respectively.
\end{cor}
Operators ${\mathcal H}^{(\alpha)}, \ \wt{\mathcal H}^{(\alpha)}$ are
the two dual Hamiltonians of the $q$-deformed
2-particle Toda chain. If $\omega_1, \ \omega_2$ are real, both
${\mathcal H}^{(\alpha)}$ and $\wt{\mathcal H}^{(\alpha)}$ are
essentially self-adjoint in the space of
smooth functions on the line which decrease faster than
$e^{-|\omega x|}$, where $ \omega = \max (\omega_1,\omega_2).$
When $\omega_1 =\ov{\omega\!}_{\,2}$, `physical' self-adjoint
Hamiltonians are ${\mathcal H}^{(\alpha)}+\wt{\mathcal H}^{(\alpha)}$
and $i({\mathcal H}^{(\alpha)}-\wt{\mathcal H}^{(\alpha)})$.

Using the explicit formulae for the Whittaker vectors
$\Phi^{(\al)}_\la, \wh\Phi^{(\al)}_\la$, we may
express Whittaker functions $w^{(\bal)}_\gamma$ in integral form:
\be\label{so1}
w^{(\bal)}_\gamma(x)=N_\gamma
e^{-\frac{\pi(\omega_1+\omega_2)x}{\omega_1\omega_2}}
\Big(\wh\Phi^{(\al_1)}_{-i\gamma-\omega_1-\omega_2}\,,
e^{-\frac{\pi x}{\omega_2}H}
\Phi^{(\al_2)}_{-i\gamma-\omega_1-\omega_2}\Big) = \\
N_\gamma\,e^{\frac{\pi i\gamma x}{\omega_1\omega_2}}\int
e^{\frac{2\pi(\omega_1+\omega_2)t}{\omega_1\omega_2}}\;
\ov{\wh\Phi^{(\al_1)}_{-i\gamma-\omega_1-\omega_2}(\ov{t}\,)}
\;\Phi^{(\al_2)}_{-i\gamma-\omega_1-\omega_2}(t+x)dt.
\ee
where one introduces the normalization factor $N_\gamma$
for the future convenience as follows:
\be
N_\gamma={\textstyle\frac{1}{\omega_1\omega_2}}\,
e^{-\frac{\pi i}{2}[B_{2,2}(i\gamma)-B_{2,2}(0)]}
\ee
with the polynomial $B_{2,2}(z)$ defined by (\ref{ds6}).

In particular, substituting in \eqref{so1} the expressions
\eqref{es3a}, \eqref{es3c}, and using \eqref{ds11} we get for
the Whittaker function $w^{(0,1)}_\gamma\equiv w^{(-)}_\gamma$
the integral representation
\be\label{es6}
w^{(-)}_\gamma(x)=N_\gamma\,
e^{\frac{\pi i\gamma x}{\omega_1\omega_2}}
\int\limits_{{\mathcal C}_-} {\cal S}^{-1}
\Big(it+{\textstyle\frac{i}{2}}\,\gamma
-{\textstyle\frac{i\omega_1\omega_2}{2\pi}}\log\zen\Big)
\,\times\\
{\cal S}^{-1}\Big(\!-it-ix-{\textstyle\frac{i}{2}}\,\gamma+
{\textstyle\frac{1}{2}}(\omega_1\!+\!\omega_2)-
{\textstyle\frac{i\omega_1\omega_2}{2\pi}}\log\zen\Big)\,
e^{\frac{2\pi i\gamma t}{\omega_1\omega_2}} \ dt,
\ee
where the contour belongs asymptotically to the sectors
\be\label{c-}
{\rm arg}\,\omega_1+\frac{\pi}{2}\ < \ {\rm arg}\,t<
\frac{1}{2}({\rm arg}\,\omega_1+{\rm arg}\,\omega_2)+\pi ,
\\
{\rm arg}\,\omega_1-\frac{\pi}{2} \ < \ {\rm arg}\,t<
\frac{1}{2}({\rm arg}\,\omega_1+{\rm arg}\,\omega_2)
\ee
(at this point one can relax the `physical' constraints
imposed on parameters $\omega_1,\,\omega_2$)
and lies between the two sets of poles of the integrand:
$$
\hspace{2cm}
t^{(-)}_{n_1,n_2}=-{\textstyle\frac{\gamma}{2}}+
{\textstyle\frac{\omega_1\omega_2}{2\pi}}\,\log\zen +
i(n_1\omega_1+n_2\omega_2)\,,\hspace{2cm}n_1,n_2\geq 0,
$$
$$
t^{(-)}_{m_1,m_2}(x)=-x-{\textstyle\frac{\gamma}{2}}-
{\textstyle\frac{\omega_1\omega_2}{2\pi}}\,\log\zen-
i\Big[(m_1\!+\!{\textstyle\frac{1}{2}})\omega_1+
(m_2\!+\!{\textstyle\frac{1}{2}})\omega_2\Big],
\ \ \ \ \ m_1,m_2\geq 0.
$$
(See \eqref{zp}, \eqref{cs} for the description  of the poles and the
zeros of ${\cal S}(y)$.) The choice of the integration
contour assures convergence and provides a natural regularization
of the divergent inner product. Indeed, to see that the  integral in
(\ref{es6}) is well defined observe that due to (\ref{as1}) the
integrand has the asymptotics
$$ e^{-\frac{\pi i t^2}{\omega_1\omega_2}+t(...)}\,.$$
But in sectors (\ref{c-}) the quadratic exponential decreases.
Hence, the integral (\ref{es6}) is absolutely convergent.

In a similar way, the function
$w^{(1,0)}_\gamma\equiv w^{(+)}_\gamma$
corresponding to (\ref{es3b}), (\ref{es3d}) admits the integral
representation
\be\label{es7}
w^{(+)}_\gamma(x)=N_\gamma\,
e^{\frac{\pi i\gamma x}{\omega_1\omega_2}}
\int\limits_{{\mathcal C}_+} {\cal S}
\Big(it+{\textstyle\frac{1}{2}}(\omega_1\!+\!\omega_2)
-{\textstyle\frac{i\omega_1\omega_2}{2\pi}}\log\zen\Big)\, \times \\
{\cal S}\Big(\!-it-ix+\omega_1\!+\!\omega_2-
{\textstyle\frac{i\omega_1\omega_2}{2\pi}}\log\zen\Big)\,
e^{\frac{2\pi i\gamma t}{\omega_1\omega_2}}\ dt,
\ee
where the contour belongs asymptotically to the sectors
\be\label{c+}
\frac{1}{2}({\rm arg}\,\omega_1+{\rm arg}\,\omega_2)+\pi<
{\rm arg}\ t<{\rm arg}\,\omega_2+\frac{3\pi}{2}, \\
\frac{1}{2}({\rm arg}\,\omega_1+{\rm arg}\,\omega_2)
<{\rm arg}\,t <{\rm arg}\,\omega_2+\frac{\pi}{2}
\ee
and lies between the two sets of poles of the integrand:
$$
\hspace{0.5cm}
t^{(+)}_{n_1,n_2}=
{\textstyle\frac{\omega_1\omega_2}{2\pi}}\,\log\zen -
i\Big[(n_1\!+\!{\textstyle\frac{1}{2}})\omega_1+
(n_2\!+\!{\textstyle\frac{1}{2}})\omega_2\Big],
\hspace{1cm}n_1,n_2\geq 0,
$$
$$
\ \ \ \ t^{(+)}_{m_1,m_2}(x)=
-x-{\textstyle\frac{\omega_1\omega_2}{2\pi}}\,\log\zen+
i(m_1\omega_1+m_2\omega_2),\ \ \ \ \ \  m_1,m_2\geq 0.
$$
The integral (\ref{es7}) is absolutely convergent.

Quite similarly, one can construct the function
$w^{(0,0)}_\gamma(x)\equiv w^{(0)}_\gamma(x)$ using the Whittaker
vectors $\wh\Phi^{(0)}_\la$ and $\Phi^{(0)}_\la$:
\be\label{es8}
w^{(0)}_\gamma(x)=
N_\gamma\,e^{\frac{\pi i\gamma x}{\omega_1\omega_2}}
\int\limits_{{\mathcal C}_0}
{\cal S}^{-1}\Big(it+{\textstyle\frac{i}{2}}\,\gamma
-{\textstyle\frac{i\omega_1\omega_2}{2\pi}}\log\zen\Big)
\,\times\\
{\cal S}\Big(\!-it-ix+\omega_1\!+\!\omega_2-
{\textstyle\frac{i\omega_1\omega_2}{2\pi}}\log\zen\Big)\,
e^{\frac{2\pi i\gamma t}{\omega_1\omega_2}}\,dt\,,
\ee
where the contour ${\cal C}_0$ belongs asymptotically to the sectors
\be
{\rm arg}\,\omega_1+\frac{\pi}{2}<
{\rm arg}\ t<\frac{1}{2}({\rm arg}\,\omega_1+{\rm arg}\,\omega_2)+
\pi,\\
\frac{1}{2}({\rm arg}\,\omega_1+{\rm arg}\,\omega_2)
<{\rm arg}\,t <{\rm arg}\,\omega_2+\frac{\pi}{2}
\ee
and lies below the poles of the integrand.

Thus, the functions (\ref{es6}), (\ref{es7}), and (\ref{es8})
are the proper solutions to corresponding spectral problems
\be\label{t1}
\hspace{-4mm}
\Big[1+q^{-1}g^{2\omega_1}e^{\frac{2\pi x}{\omega_2}}\Big]
w_\gamma^{(-)}(x-i\omega_1)
+w^{(-)}_\gamma(x+i\omega_1)=\Big(e^{\frac{\pi\gamma}{\omega_2}}+
e^{-\frac{\pi\gamma}{\omega_2}}\Big)w^{(-)}_\gamma(x)\,,
\ee
\be\label{t2}
w^{(+)}_\gamma(x-i\omega_1)+
\Big[1+qg^{2\omega_1}e^{\frac{2\pi x}{\omega_2}}\Big]
w_\gamma^{(+)}(x+i\omega_1)
=\Big(e^{\frac{\pi\gamma}{\omega_2}}+
e^{-\frac{\pi\gamma}{\omega_2}}\Big)w^{(+)}_\gamma(x)\,,
\ee
\be\label{t3}
\hspace{-2mm}
w_\gamma^{(0)}(x-i\omega_1)+w^{(0)}_\gamma(x+i\omega_1)+
g^{2\omega_1}e^{\frac{2\pi x}{\omega_2}}w^{(0)}_\gamma(x)
=\Big(e^{\frac{\pi\gamma}{\omega_2}}+
e^{-\frac{\pi\gamma}{\omega_2}}\Big)w^{(0)}_\gamma(x)\,.
\ee
Besides, these solutions are the eigenfunctions for the dual spectral
problems where $\omega_1\leftrightarrow\omega_2$.

The solutions $w^{(\pm)}_\gamma$ described above appear to be close
to the $q$-Macdonald functions of the first and second kind
which arise in the context of relativistic Toda chain \cite{OR}.
However, the deformations of the Macdonald function
have been investigated in the framework of the standard $q$-analysis
\cite{GR} for the typical region $|q|<1$ (which evidently fails in
the case $|q|=1$) and without any reference to the dual symmetry.

Formulae \eqref{so1} will be referred to as the
Gauss-Euler representation for Whittaker functions. The integral
representations for $w^{(\e)}_\gamma(x),\, (\e=0,\pm 1)$ are
the degenerations of more general $q$-hypergeometric function
\cite{NU}.

We shall see later that the technique of QISM yields a different
integral representation for Whittaker functions which is a
$q$-deformation of the Mellin-Barnes integrals.

\subsection{Analytic  properties}
Let us give the summary of the analytic properties of the
Whittaker functions which may be derived directly from the
Gauss-Euler representation.
\begin{lem}
$w_\gamma^{(\pm)}$ and $w^{(0)}_\gamma$
can be extended to the entire functions  in
$\gamma\in\CC$. As a  function of $x\in\CC$, $w_\gamma^{(-)}(x)$
has   poles   at
\be
x=-{\textstyle\frac{\omega_1\omega_2}{\pi}}\,\log\zen
-i(k_1+{\textstyle\frac{1}{2}})\omega_1-
i(k_2+{\textstyle\frac{1}{2}})\omega_2, \ \ \ \ \ k_1, k_2\geq 0.
\ee
Similarly, the function $w_\gamma^{(+)}(x)$ has poles at
\be
x=-{\textstyle\frac{\omega_1\omega_2}{\pi}}\,\log\zen+
i(k_1+{\textstyle\frac{1}{2}})\omega_1+
i(k_2+{\textstyle\frac{1}{2}})\omega_2,\ \ \ \ \ k_1, k_2\geq 0.
\ee
The function $w^{(0)}_\gamma(x)$ is an entire one in $x\in\CC$.
\end{lem}
\begin{lem}
\begin{align}
w_\gamma^{(+)}(x)&= {\cal S}\Big(-ix+{\textstyle\frac{1}{2}}
(\omega_1\!+\!\omega_2)-
{\textstyle\frac{i\omega_1\omega_2}{\pi}}\,\log\zen\Big)\;
w_\gamma^{(-)}(x).
\end{align}
\end{lem}
\begin{lem}
For any $\gamma\in\CC$ such that
\begin{align}\label{ch0}
{\rm arg}\,\gamma \notin[{\rm
arg}\,\omega_2-{\textstyle\frac{\pi}{2}}, {\rm arg}\,\omega_1-
\textstyle{\frac{\pi}{2}}]\;\bigcup\; [{\rm arg}\,
\omega_2+\frac{\pi}{2},
{\rm arg}\,\omega_1+{\textstyle\frac{\pi}{2}}]
\end{align}
the following asymptotics holds as $x$ tends to infinity in
the sector
\begin{align}
\label{ch4} {\rm arg}\,\omega_1+\frac{\pi}{2}<{\rm arg}\,x<
{\rm arg}\,\omega_2+\frac{3\pi}{2}:
\end{align}
\begin{align}
\label{ch3} w^{(\e)}_\gamma(x) &= c(\gamma)\, e^{\frac{\pi
i\gamma x}{\omega_1\omega_2}}\Big[1+o(1)\Big]+ c(-\gamma)\,
e^{-\frac{\pi i\gamma x}{\omega_1\omega_2}} \Big[1+o(1)\Big].
\end{align}
$(\e=0,\pm)$, where the function $c(\gamma)$ is defined
by (\ref{ch1}).
\end{lem}

We shall call $c(\gamma)$ the quantum Harish-Chandra function
associated with $U_q(\mathfrak{sl}(2,\RR))$.

\subsection{Mellin-Barnes representation}
To make a comparison
with the formulae provided by the Quantum Inverse Scattering Method
we need a different integral representation of the Whittaker
functions. Put
\be\label{mb1}
\psi^{(\e)}_\gamma(x)=e^{\frac{\pi i\gamma x}{\omega_1\omega_2}}
\int\limits_{{\cal C}'_e}
c(\zeta)c(\zeta\!+\!\gamma)\,
e^{-\frac{\pi i\e}{\omega_1\omega_2}[\zeta^2+\gamma\zeta]}
\;e^{\frac{2\pi i\zeta x}{\omega_1\omega_2}}\,d\zeta
\ee
where the contour ${\cal C}'_\e$ is above the poles of the
integrand and belongs in the left (right) half-plane in
$\zeta\in\CC$ to the sectors where the exponential
$e^{-\frac{\pi(\e-1)\zeta^2}{\omega_1\omega_2}}$
($e^{-\frac{\pi(\e+1)\zeta^2}{\omega_1\omega_2}}$) is quadratically
vanish. The integral (\ref{mb1}) is absolutely convergent for
any $x\in\CC$ provided $\e\neq\pm 1$.
In degenerate case $\e=-1$ the integral is convergent provided
the condition
\be\label{rest1'}
{\rm arg}\,x\notin \Big[{\rm arg}\,\omega_2-\frac{\pi}{2},
{\rm arg}\,\omega_1-\frac{\pi}{2}\Big]\,,
\ee
while for $\e=1$ it is defined in the region
\be
\label{rest2'}
{\rm arg}\,x\notin \Big[{\rm arg}\,\omega_2+\frac{\pi}{2},
{\rm arg}\,\omega_1+\frac{\pi}{2}\Big]\,.
\ee
Using the properties of double sine it can be directly
verified that the function $\psi^{(\e)}_\gamma(x)$
satisfies to equations (\ref{gt1}) where $\al_1-\al_2=\e$.

\begin{prop}
For $\e=0,\pm 1$
\be\label{mb3}
w_\gamma^{(\e)}(x)=\psi_\gamma^{(\e)}(x)\,.
\ee

\end{prop}

The expression (\ref{mb1}) will be referred to as the
($q$-deformed) Mellin-Barnes representation for Whittaker functions.
It will be shown below that this is the very representation which can
be easily generalized to those for $N$-particle $q$-deformed
Toda chain.

\subsection{Limit to $SL(2,\RR)$ Toda chain}\label{clas}
Let $\omega_k>0\,,\;(k=1,2)$. Suppose that the "coupling constant"
$\zen(\bo)$ has the asymptotics such that
\be\label{cl1}
\hspace{1cm}
\zen^{\omega_1}(\bo)=\frac{2\pi}{\omega_2}\,[1+O(\omega_2^{-1})\,]
\hspace{2cm}(\omega_2\to\infty)
\ee
For example, the simplest (and standard) choice
$\zen^{\omega_1}(\bo)=\frac{q-q^{-1}}{i\omega_1}$ satisfies this
condition. After the rescaling $x\to\frac{\omega_2}{\pi}\,x$ the
equations (\ref{t1}), (\ref{t2}), and (\ref{t3}) take the form
\bse\label{cl2}
\be\label{cl2a}
\hspace{-2mm}
\Big\{\Big[1+q^{-1}\zen^{2\omega_1}e^{2x}\Big]
e^{-\frac{\pi i\omega_1}{\omega_2}\d_{x}}+
e^{\frac{\pi i\omega_1}{\omega_2}\d_x}\Big\}w^{(-)}_\gamma(x)
=\Big(e^{\frac{\pi\gamma}{\omega_2}}\!+\!
e^{-\frac{\pi\gamma}{\omega_2}}\Big)w^{(-)}_\gamma(x)\;,
\ee
\be\label{cl2b}
\hspace{-2mm}
\Big\{e^{-\frac{\pi i\omega_1}{\omega_2}\d_{x}}+
\Big[1+q^{-1}\zen^{2\omega_1}e^{2x}\Big]
e^{\frac{\pi i\omega_1}{\omega_2}\d_x}\Big\}w^{(+)}_\gamma(x)
=\Big(e^{\frac{\pi\gamma}{\omega_2}}\!+\!
e^{-\frac{\pi\gamma}{\omega_2}}\Big)w^{(+)}_\gamma(x)\;,
\ee
\be\label{cl2c}
\hspace{-2mm}
\Big\{e^{-\frac{\pi i\omega_1}{\omega_2}\d_{x}}+
e^{-\frac{\pi i\omega_1}{\omega_2}\d_{x}}+g^{2\omega_1}e^{2x}\Big\}
w^{(0)}_\gamma(x)
=\Big(e^{\frac{\pi\gamma}{\omega_2}}\!+\!
e^{-\frac{\pi\gamma}{\omega_2}}\Big)w^{(0)}_\gamma(x)\;,
\ee
\ese
In the limit (\ref{cl1}) the equations (\ref{cl2}) are reduced to
$SL(2,\RR)$ Toda equation
\be\label{cl3}
\Big\{p^2+4e^{2x}\Big\}w_\gamma(x)=\gamma^2w_\gamma(x)\;.
\ee
where $p=-i\omega_1\d_x$ and $\omega_1$ plays the role of Planck
constant. Note that the more general equation (\ref{gt1a}) has the
same limit (\ref{cl3}). The solution to (\ref{cl3}) with appropriate
asymptotic behavior is written in terms of Macdonald function
\be\label{cl4}
w_\gamma(x)=K_{\frac{\gamma}{i\omega_1}}
({\textstyle\frac{2}{\omega_1}}\,e^x)\,.
\ee
\begin{lem}
\be\label{cl5}
\lim_{\omega_2\to\infty}\psi^{(\e)}
({\textstyle\frac{\omega_2}{\pi}}\,x)\,=\,
\frac{1}{\pi\omega_1}\,
K_{\frac{\gamma}{i\omega_1}}
({\textstyle\frac{2}{\omega_1}}\,e^x)\,.
\ee
\end{lem}
{\bf Proof.} Using the formula
\be\label{cl6}
\lim_{\omega_2\to\infty}\,\sqrt{2\pi}
\left(\frac{2\pi\omega_1}{\omega_2}\right)^{\frac{\scriptstyle 1}
{\scriptstyle 2}-\frac{\scriptstyle z}{\scriptstyle\omega_1}}
S^{-1}_2(z)\,=\,\Gamma\Big(\frac{z}{\omega_1}\Big)\,,
\ee
proved in \cite{Rui2}, one easily finds that the quantum
Harish-Chandra function (\ref{ch1})
reduces to the usual $\Gamma$-function:
\be\label{cl7}
\lim_{\omega_2\to\infty}c(\zeta)=
{\textstyle\frac{1}{2\pi\omega_1}}\,
\omega_1^{\frac{\zeta}{i\omega_1}}
\Gamma\Big(\frac{\zeta}{i\omega_1}\Big)\,,
\ee
provided that asymptotics (\ref{cl1}) holds. (This function is
closely related to the standard Harish-Chandra function for the Toda
chain \cite{STS-toda}; the difference with the usual definition
is due to a different normalization of solutions).
Hence, in the limit $\omega_2\to\infty$ the rescaled function
(\ref{mb1}) acquires the form
\be
\lim_{\omega_2\to\infty}\psi^{(\e)}
({\textstyle\frac{\omega_2}{\pi}}\,x)\,=\\ =\,
\frac{1}{\pi\omega_1}\,
\left\{\frac{1}{4\pi\omega_1}\,
\Big(\,\frac{e^x}{\omega_1}\,\Big)^{\frac{i\gamma}{\omega_1}}
\int\Gamma\Big(\frac{\zeta}{i\omega_1}\Big)
\Gamma\Big(\frac{\zeta+\gamma}{i\omega_1}\Big)\,
\Big(\,\frac{e^x}{\omega_1}\,\Big)^{\frac{2i\zeta}{\omega_1}}
d\zeta\right\}\,,
\ee
where the contour is parallel to real axis and above the poles of the
integrand. The expression in brackets is exactly the Macdonald
function (\ref{cl4}) in Mellin-Barnes representation.\square

\section{$N$-particle $q$-Toda chain and duality}
The extension of the formalism described above to the case of
$N$-particle Toda chain may be performed directly with the help
of the `free field representation' for $U_q(\mathfrak{sl}(N,\RR))$,
i.e., the homomorphism  of $U_q(\mathfrak{sl}(N,\RR))$ into an
appropriate multidimensional quantum torus. Instead, we shall
describe a different approach based on `lattice Lax representation
with spectral parameter'. As usual,  Lax representation
allows to construct quantum Hamiltonians for a bunch
of related systems: periodic Toda chain, open Toda chain,
as well as different degenerate systems obtained by removing some of
the potential terms from the Hamiltonians. Of course, the
choice of the model in question depends on our choice of the quantum
$R$-matrix. The obvious choice  is the standard trigonometric
$4\times 4$ $R$-matrix; to get more freedom in the choice of the
model we may use {\em twisted trigonometric $R$-matrices\/}.
In all cases, there is a natural homomorphism of the corresponding
quantum algebra into the tensor product of noncommutative tori; this
allows to introduce the corresponding dual system realized by
means of the natural representation of the product of modular
dual quantum tori in the same Hilbert space.
The entire picture of modular duality is thus fully
generalized to the $N$-particle case. We would like to point out
that in
$R$-matrix formalism it is more convenient to work with
$U_q(\mathfrak{gl}(N,\RR))$ and reduce the final formulae to the
case of $U_q(\mathfrak{sl}(N,\RR))$ by the standard way.

The main advantage provided by the use of the lattice Lax
representation is the possibility to get inductive integral
representations for the wave functions in question and generalization
of the above construction to the periodic case as it was done in
\cite{KL3}.

\subsection{The models}
$q$-Toda chain, or relativistic Toda chain (RTC),
was introduced by Ruijsenaars \cite{Rui}.
The periodic chain can be described by the Hamiltonian
\be\label{trm4a'}
H_1(x_1,p_1;\ldots;x_{\N},p_{\N})=\sum_{n=1}^N
\Big\{1+q^{-1}\zen^{2\omega_1}e^{\frac{2\pi}{\omega_2}
(x_n-x_{n+1})}\Big\}\,e^{\omega_1p_n}
\ee
where $x_n,\,p_n$ are the canonical coordinates and momenta with
standard commutation relations $[x_n,p_m]=i\delta_{nm}$
and the boundary condition $x_{{\N}+1}=x_1$ is imposed.
The system has exactly $N$ mutually commuting Hamiltonians
(the polynomial functions of the Weyl variables
$u^{\pm 1}_n=e^{\pm\frac{2\pi x_n}{\omega_2}}\,,
v_n=e^{\omega_1 p_n}$)
\footnote{
Higher Hamiltonians will be described below using the standard
Lax formalism.
}
.

Guided by the notion of the modular double considered above,
one can define the {\it dual} system which is determined by the
Hamiltonian
\be\label{trm4b'}
\wt H_1(x_1,p_1;\ldots;x_{\N},p_{\N})=\sum_{n=1}^N
\Big\{1+q^{-1}\zen^{2\omega_2}e^{\frac{2\pi}{\omega_1}
(x_n-x_{n+1})}\Big\}\,e^{\omega_2 p_n}
\ee
with the same boundary condition. It is evident that the systems
mutually commute.

Analogously, the open relativistic Toda chain and its dual system
are defined by the Hamiltonians
\be\label{lm10a'}
h_1(x_1,p_1;\ldots;x_{\N},p_{\N})=\sum_{n=1}^N
\Big\{1+q^{-1}\zen^{2\omega_1}e^{\frac{2\pi}{\omega_2}
(x_n-x_{n+1})}\Big\}\,e^{{\omega_1}p_n}
\ee
and
\be\label{lm10b'}
\wt h_1(x_1,p_1;\ldots;x_{\N},p_{\N})=\sum_{n=1}^N
\Big\{1+q^{-1}\zen^{2\omega_2}e^{\frac{2\pi}{\omega_1}
(x_n-x_{n+1})}\Big\}\,e^{{\omega_1}p_n}
\ee
respectively with the boundary condition $x_{{\N}+1}\equiv\infty$.
Similarly to the periodic case, each open system possesses exactly
$N$ mutually commuting Hamiltonians. Moreover, the Hamiltonians
of the dual system commute with those of original one.

The basic goal of the present section is to construct the explicit
integral representation of the common eigenfunctions for all
Hamiltonians in the case of the {\it open} $N$-particle RTC.
This will be done in the framework of the QISM approach for the
{\it periodic} RTC.

\subsection{Twisted trigonometric R-matrix}
In order to investigate the relativistic Toda chain using the
quantum version of the corresponding classical Lax matrix
\cite{Sur}, one needs to introduce the notion of the twisted
$R$-matrix \cite{reshetik}. Let
\be\label{tm0}
R(z/w)\;=\;\frac{1}{z^2\!-\!w^2}\;\times\\ \times
\left(\begin{array}{cccc}qz^2\!-\!q^{-1}w^2&0&0&0\\
0&z^2\!-\!w^2 & (q\!-\!q^{-1})zw&0\\
0&(q\!-\!q^{-1})zw& z^2\!-\!w^2 & 0\\
0&0&0&qz^2\!-\!q^{-1}w^2\end{array}\right)
\ee
be the $R$-matrix in principal gradation satisfying the standard
Yang-Baxter equation. Consider the twisting of $R$-matrix
(\ref{tm0}):
\be\label{tm1}
R_\theta(z/w)\,=\,F_{21}(\theta)R(z/w)F^{-1}_{12}(\theta)
\ee
with
\be\label{tm2}
F_{12}(\theta)\;\equiv\;F^{-1}_{21}(\theta)\,=\,
\exp\Big\{\frac{\theta}{4}\Big(1\otimes\sigma_3-
\sigma_3\otimes 1\Big)\Big\}
\ee
where $\sigma_3$ is the Pauli matrix. One gets
\be\label{tm3}
R_\theta(z/w)\;=\;\frac{1}{z^2\!-\!w^2}\,
\left(\begin{array}{cccc}
a(z,w)&0&0&0\\
0&b(z,w)&c(z,w)&0\\
0&c(z,w)&\ov{b}(z,w)&0\\
0&0&0&a(z,w)
\end{array}\right)
\ee
where
\be\label{tm4}
\begin{array}{l}
a(z,w)\,=\,qz^2-q^{-1}w^2\\
b(z,w)\,=\,e^\theta(z^2-w^2)\\
\ov{b}(z,w)\,=\,e^{-\theta}(z^2-w^2)\\
c(z,w)\,=\,(q-q^{-1})zw
\end{array}
\ee
It is easy to verify that $R_\theta(z/w)$ satisfies the same
Yang-Baxter equation as $R(z/w)$.

A quantum Lax operator $L(z)$ is, by definition,
a $2\times2$-matrix
\be\label{main}
L(z)\,=\,\left(
\begin{array}{ll}
L^{11}(z) & L^{12}(z)\\
L^{21}(z) & L^{22}(z)
\end{array}\right)
\ee
with operator-valued entries which satisfies the fundamental
commutation relations
\be\label{rll}
R_\theta(z/w)L(z)\otimes L(w)\,=\,
(1\otimes L(w))(L(z)\otimes 1)R_\theta(z/w)\,.
\ee
We define the quantum determinant of the matrix (\ref{main}) by the
formula
\be\label{qd}
{\rm det}_q\,L(z)=L^{11}(zq^{1/2})L^{22}(zq^{-1/2})-e^\theta
L^{12}(zq^{1/2})L^{21}(zq^{-1/2})\,.
\ee

\subsection{Lax operator and monodromy matrix}
As usual in the Quantum Inverse Scattering Method, the entries of
the quantum Lax operator generate the basic Hopf algebra
${\cal A}_R$ (defined implicitly by the fundamental commutation
relation (\ref{rll}) which underlies all the associated
quantum integrable systems; to get a particular system, we need to
fix its representation. The representation which yields the
$q$-deformed Toda chain is provided by the following construction.

Let $\omega_1, \omega_2 \in\CC$. We consider a lattice system
with local quantum Lax operators
\be\label{lm1m} L_n(z)\,
=\,\left( \begin{array}{cc}
z-z^{-1}e^{\omega_1 p_n}\ \ \ &
\zen^{\omega_1}\,e^{-\frac{2\pi{\scriptstyle x_n}}{\omega_2}}\\
-\zen^{\omega_1}\,e^{\frac{2\pi{\scriptstyle x_n}}{\omega_2}+
\omega_1 p_n} & 0
\end{array}\right)
\ee
where $x_n,\,p_n$ are the canonical coordinates and momenta with
the commutation relations $[x_n,p_m]=i\delta_{nm}$ and
$\zen$ is a real parameter (possibly depending
on $\bo$). On the classical level the Lax matrices (\ref{lm1m})
have been introduced in \cite{Sur}.
\begin{prop}
Lax operator (\ref{lm1m}) satisfies the commutation relations
(\ref{rll}) with the quantum $R$-matrix (\ref{tm3}),  (\ref{tm4}),
where
\be\label{lm3}
q=e^{i\pi\frac{\omega_1}{\omega_2}}\,,
\ee
and
\be\label{lm3b}
e^\theta=q.
\ee
\end{prop}
The monodromy matrix for the $N$-{\it periodic} chain is defined
in the standard way:
\be\label{lm5}
T_{\N}(z)=L_{\N}(z)\ldots L_1(z)\equiv\left
(\begin{array}{cc}
A_{\N}(z)&B_{\N}(z)\\
C_{\N}(z) & D_{\N}(z)
\end{array}\right)\,.
\ee
By the usual Hopf algebra properties, the  entries of $T(z)$
satisfy the same commutation relations as the corresponding
entries of the Lax operators.

The quantum determinant of the Lax operator (\ref{lm1m}) is
\be\label{lm4}
{\rm det}_qL(z)=\zen^{2\omega_1}e^{\omega_1{p_n}}\,.
\ee
It is simple to show that the quantum determinant of monodromy
matrix
\be\label{lm6}
\mbox{det}_q T_{\N}(z)=A_{\N}(zq^{1/2})D_{\N}(zq^{-1/2})-
qB_{\N}(zq^{1/2})C_{\N}(zq^{-1/2})
\ee
obeys the property
\be
\mbox{\rm det}_q T_{\N}(z)=\mbox{\rm det}_q
L_{\N}(z)\cdot\ldots\cdot \mbox{\rm det}_q L_1(z)\,.
\ee
Hence, due to (\ref{lm4}),
\be\label{lm7}
{\rm det}_qT_{\N}(z)=\zen^{2N\omega_1}
\prod_{n=1}^{N}e^{\omega_1{p_n}}\,.
\ee
Note that in the twisted case the quantum determinant is no longer
a central element in the quantum algebra.

Following the same line of argument as in section 1, we may
introduce the modular dual system by
\be\label{lm8}
\wt L_n(z)\,=\,\left( \begin{array}{cc}
z-z^{-1}e^{\omega_2p_n}\ \ \ &
\zen^{\omega_2}\,e^{-\frac{2\pi{\scriptstyle x_n}}{\omega_1}}\\
-\zen^{\omega_2}\,e^{\frac{2\pi{\scriptstyle x_n}}{\omega_1}+
\omega_2p_n} & 0
\end{array}\right)\,.
\ee
The operator (\ref{lm8}) satisfies the commutation relation
(\ref{rll}) with the twisted $R$-matrix (\ref{tm3}), (\ref{tm4})
with the only change $q\to\wt q$, $\theta\to\wt \theta$ where
$\wt q=e^{\wt \theta}=e^{i\pi\frac{\omega_2}{\omega_1}}$.
The dual monodromy matrix is defined  by
\be
\wt
T_{\N}(z)=\wt L_{\N}(z)\ldots \wt
L_1(z)\equiv\left(\begin{array}{cc} \wt A_{\N}(z)&\wt B_{\N}(z)\\
\wt C_{\N}(z) & \wt D_{\N}(z)\end{array}\right).
\ee
The system describing by the Lax operators (\ref{lm1m}), (\ref{lm8})
may be referred to as the modular relativistic Toda chain.

\subsection{Hamiltonians}
As usual, the transfer matrix
\be\label{trm0}
t_{\N}(z)=A_{\N}(z)+D_{\N}(z)
\ee
satisfies to commutation relations
\be\label{trm1}
[t_{\N}(z),t_{\N}(w)]=0\,.
\ee
The same is true for the dual transfer matrix
\be\label{trm1'}
\wt t_{\N}(z)=\wt A_{\N}(z)+\wt D_{\N}(z);
\ee
moreover, the modular duality implies that
\be\label{tr}
[t_{\N}(z),\wt t_{\N}(w)] = 0.
\ee
Clearly, $t_{\N}(z)$ has the following structure:
\be\label{trm2}
t_{\N}(z)=\sum_{k=0}^N(-1)^kz^{N-2k}
H_k(x_1,p_1;\ldots;x_{\N},p_{\N})
\ee
where $H_0=1$ and
\be\label{trm3b}
H_{\N}(p_1,\ldots,p_{\N})=
\exp\left\{\omega_1\sum_{n=1}^Np_n\right\}\,,
\ee
\be\label{trm4a}
H_1(x_1,p_1;\ldots;x_{\N},p_{\N})=\sum_{n=1}^N
\Big\{1+q^{-1}\zen^{2\omega_1}e^{\frac{2\pi}{\omega_2}
(x_n-x_{n+1})}\Big\}\,e^{\omega_1 p_n}\,,
\ee
\be\label{trm4b}
H_{{\N}-1}(x_1,p_1;\ldots;x_{\N},p_{\N})=H_{\N}\sum_{n=1}^N
\Big\{1+q^{-1}\zen^{2\omega_1}e^{\frac{2\pi}{\omega_2}
(x_{n-1}-x_n)}\Big\}\,e^{-\omega_1 p_n}\,,
\ee
where in (\ref{trm4a}), (\ref{trm4b}) the periodicity is assumed:
$x_{{\N}+1}=x_1$.
Hence, due to (\ref{trm1}) the {\it periodic} RTC has exactly $N$
commuting operators.

The following statement is true: the operator $A_{\N}(z)$ is
the generating function for the Hamiltonians of $N$-particle
{\it open} RTC:
\be\label{lm8'}
A_{\N}(z)=\sum_{k=0}^N(-1)^kz^{N-2k}
h_k(x_1,p_1;\ldots;x_{\N},p_{\N})
\ee
where $h_0=1$ and
\be\label{lm9b}
h_{\N}(p_1,\ldots,p_{\N})=
\exp\left\{\omega_1\sum_{n=1}^Np_n\right\}\,,
\ee
\be\label{ho1}
h_1(x_1,p_1;\ldots;x_{\N},p_{\N})=\sum_{n=1}^N
\Big\{1+q^{-1}\zen^{2\omega_1}e^{\frac{2\pi}{\omega_2}
(x_n-x_{n+1})}\Big\}\,e^{\omega_{1}p_n}\,,
\ee
\be\label{ho2}
h_{{\N}-1}(x_1,p_1;\ldots;x_{\N},p_{\N})=h_{\N}\sum_{n=1}^N
\Big\{1+q^{-1}\zen^{2\omega_1}e^{\frac{2\pi}{\omega_2}
(x_{n-1}-x_n})\Big\}\,e^{-\omega_1 p_n}
\ee
assuming $x_{{\N}+1}\equiv\infty$ in (\ref{ho1}) and
$x_0\equiv-\infty$ in (\ref{ho2}).

The second set of the Hamiltonians $\wt H_1,\dots \wt H_{\N}$ and
$\wt h_1,\dots \wt h_{\N}$ are obtained from the former one by
the flip $\omega_1\leftrightarrow\omega_2$.
\begin{lem}
\begin{enumerate}
\item
Suppose that $\omega_1,\omega_2$ are real; then all
coefficients of $t_{\N}(z)$, $\wt t_{\N}(z)$ and
$A_{\N}(z),\wt A_{\N}(z)$
are formally self-adjoint in $L_2(\RR^N)$.
\item
Suppose that ${\rm Im}\,\omega_1\neq 0 $ and
$\ov\omega_1=\omega_2$; then all coefficients
of $t_{\N}(z),\;\wt t_{\N}(z)$ and $A_{\N}(z),\;\wt A_{\N}(z)$
are normal operators and their `real' and
'imaginary' parts ($X+\wt X$, $i(X-\wt X)$)
are formally self-adjoint.
\end{enumerate}
\end{lem}

\subsection{Integral representation for the wave functions:
inductive procedure}
Our goal is to get an inductive integral representation
for the wave functions of the multiparticle open
relativistic Toda chain. The approach described below is an
analytical version of the algebraic method of separation of
variables invented by Sklyanin \cite{Skl1}.

Set $\bgamma =(\gamma_1,\dots ,\gamma_{{\N}-1})\in\RR^{N-1},
\,\bfit x=(x_1, \dots ,x_{{\N}-1})\in\RR^{N-1}$. Let
$\psi_\bgamma(\bfit x)$
be the common wave function for the dual open RTC systems with
$N-1$ particles:
\be\label{e1}
A_{{\N}-1}(z)\psi_\bgamma(\bfit x)=\prod_{m=1}^{N-1}
\Big(z-z^{-1}e^{\frac{2\pi\gamma_m}{\omega_2}}\Big)\;
\psi_\bgamma(\bfit x)\,,
\ee
\be\label{e1tilde}
\wt A_{{\N}-1}(z)\psi_\bgamma(\bfit x)=\prod_{m=1}^{N-1}
\Big(z-z^{-1}e^{\frac{2\pi\gamma_m}{\omega_1}}\Big)\;
\psi_\bgamma(\bfit x).
\ee
The key point of the inductive procedure (described for
the first time in \cite{KL2} for the ordinary Toda chain) is
to compute the action on $\psi_\bgamma(\bfit x)$
of the $N$-particle Hamiltonians. It turns out that such an action
`preserves' the form of wave function
\footnote{
This idea goes back to M. Gutzviller \cite{Gu} who explicitly
calculated such an action on the 2 and 3-particle eigenfunctions
for the Toda chain.
}.
This computation, which starts
with the case $N=2$, is based on the ordinary $RTT$ commutation
relations for the quantum monodromy matrix; the inductive formula
given below is based on a self-consistent choice
of the normalization for the wave functions.
\begin{prop}\label{AN}
There exists the unique solution
$\psi_{\gamma_{1},\ldots,\gamma_{{N}-1}}(x_1,\ldots,x_{{\N}-1})$
to the common spectral problem (\ref{e1}), (\ref{e1tilde}) such that
for any $N\ge 2$ the eigenfunction $\psi_\bgamma$ is an entire
function in $\bgamma\in\CC^{N-1}$ satisfying to relations
\be\label{af1a}
A_{\N}\Big(e^{\frac{2\pi\gamma_j}{\omega_2}}\Big)
\psi_\bgamma(\bfit x)=q^{-1} (-i\zen^{\omega_1})^N
\exp\left\{\frac{2\pi}{\omega_2}\sum_{m=1}^{N-1}\gamma_m\right\}\;
e^{-\frac{2\pi x_N}{\omega_2}}\psi_{\bgamma-i\omega_1\bfit e_j}
(\bfit x),
\ee
\be\label{af1a'}
\hspace{-2mm}
\wt A_{\N}\Big(e^{\frac{2\pi\gamma_j}{\omega_1}}\Big)
\psi_\bgamma(\bfit x)=\wt q^{\,-1}(-i\zen^{\omega_2})^N
\exp\left\{\frac{2\pi}{\omega_1}\sum_{m=1}^{N-1}\gamma_m\right\}\;
e^{-\frac{2\pi x_N}{\omega_1}}\psi_{\bgamma-i\omega_2\bfit e_j}
(\bfit x),
\ee
where $\{\bfit e_j\}$ be the standard basis of $\RR^{N-1}$.
\end{prop}
As a comment to the proposition above we remark that
$C_{\N}(z)=-q^{-1}\zen^{\omega_1}e^{\frac{2\pi x_N}{\omega_2}}
e^{\omega_1 p_N}A_{{\N}-1}(z)$. Hence, the compatibility of
(\ref{e1}) and (\ref{af1a}) follows from the quadratic $RTT$
relations; the argument for the dual system is completely similar.

Assuming that such a function is known, the heuristic idea behind
the inductive integral representation for the $N$-particle wave
function $\psi_{\la_1,\ldots,\la_{\N}}(x_1,\ldots,x_{\N})$
is to represent it as the generalized Fourier transform with
respect to the $N-1$-particle wave function
$\psi_{\bgamma}(\bfit x)$. Note that the "one-particle" solution is
$\psi_{\gamma_1}(x_1) =
e^{\frac{2\pi i\gamma_1x_1}{\omega_1\omega_2}}$; it is easy to
verify directly that this trivial function satisfies the conditions
of proposition \ref{AN}, which forms the induction basis. The exact
statement is given by theorem \ref{main1} below; here we
represent the essential ideas how to arrive to this statement.

Introduce an auxiliary wave function by
\be\label{af2}
\Psi_{\bgamma,\varepsilon}(x_1,\ldots,x_{\N})\;
\stackrel{\mbox{\tiny def}}{=}\\
\exp\left\{\frac{\pi i}{\omega_1\omega_2}\left[
\Big(\sum_{m=1}^{N-1}\gamma_m\Big)^2-
\varepsilon\sum_{m=1}^{N-1}\gamma_m\right]\right\}
e^{\raise2pt\hbox{$\frac{2\pi i}{\omega_1 \omega_2}
\scriptstyle\big(\varepsilon\,-\!\sum\limits_{m=1}^{N-1}\gamma_m\big)
\raise1pt\hbox{$\scriptstyle x_N $}$}}\,
\psi_{\bfit\gamma}(x_1,\ldots,x_{{\N}-1})\,.\hspace{-1cm}
\ee
Generalizing proposition \ref{AN}, one can prove the following
result.
\begin{prop}
The action of $A_{\N}(z),\,\wt A_{\N}(z)$ on the auxiliary wave
function (\ref{af2}) is given by
\be\label{in4}
A_{\N}(z)\Psi_{\bgamma,\varepsilon}=\left(z-z^{-1}
e^{\raise2pt\hbox{$\scriptstyle\frac{2\pi}{\omega_2}\,
\big(\varepsilon\,-\!\sum\limits_{m=1}^{N-1}\gamma_m\big)$}}\right)
\prod_{j=1}^{N-1}\Big(z-z^{-1}e^{\frac{2\pi\gamma_j}{\omega_2}}\Big)
\;\Psi_{\bgamma,\varepsilon}\;+\\+\,(-i\zen^{\omega_1})^N
e^{\frac{\pi\varepsilon}{\omega_2}}
\sum_{j=1}^{N-1}\Psi_{\bgamma-i\omega_1\bfit e_j,\varepsilon}
\prod_{s\neq j}\frac{z-e^{\frac{2\pi\gamma_s}{\omega_2}}z^{-1}}
{e^{\frac{\pi\gamma_j}{\omega_2}}-e^{\frac{2\pi\gamma_s}{\omega_2}}
e^{-\frac{\pi\gamma_j}{\omega_2}}}\;,
\ee
\be\label{in4'}
\wt A_{\N}(z)\Psi_{\bgamma,\varepsilon}=\left(z-z^{-1}
e^{\raise2pt\hbox{$\scriptstyle\frac{2\pi}{\omega_1}\,
\big(\varepsilon\,-\!\sum\limits_{m=1}^{N-1}\gamma_m\big)$}}\right)
\prod_{j=1}^{N-1}\Big(z-z^{-1}e^{\frac{2\pi\gamma_j}{\omega_1}}\Big)
\;\Psi_{\bgamma,\varepsilon}\;+\\+\,(-i\zen^{\omega_2})^N
e^{\frac{\pi\varepsilon}{\omega_1}}
\sum_{j=1}^{N-1}\Psi_{\bgamma-i\omega_2\bfit e_j,\varepsilon}
\prod_{s\neq j}\frac{z-e^{\frac{2\pi\gamma_s}{\omega_1}}z^{-1}}
{e^{\frac{\pi\gamma_j}{\omega_1}}-e^{\frac{2\pi\gamma_s}{\omega_1}}
e^{-\frac{\pi\gamma_j}{\omega_1}}}\;.
\ee
\end{prop}
Let us write formally
\be\label{rr41}
\psi_{\la_1,\ldots,\la_{\N}}(x_1,\ldots,x_{\N})=
\int\mu(\bgamma){\cal Q}(\bgamma|\blambda)
\Psi_{\bgamma,\la_1+\ldots+\la_{\N}}\,d\bgamma .
\ee
where
\be\label{meas2}
\mu(\bgamma)\;\stackrel{\mbox{\tiny def}}{=}\;
\prod_{j<k}\left\{4\omega_1\omega_2\,\sinh\frac{\pi}
{\omega_1}(\gamma_j\!-\!\gamma_k)\cdot
\sinh\frac{\pi}{\omega_2}(\gamma_j\!-\!\gamma_k)\right\}.
\ee
and ${\cal Q}(\bgamma|\blambda)$ is an unknown kernel.
The form of the integrand in (\ref{rr41}) is a natural
generalization of the one for the usual Toda chain \cite{KL2};
the latter case actually corresponds to the limit
$\omega_2\to\infty$ of RTC model, and in this
limit the function $\mu(\bgamma)$ reduces to the Sklyanin measure
\cite{Skl1}.

By assumption, $\psi_{\bgamma}(\bfit x)$ satisfies, in particular,
the equations (\ref{e1}) and (\ref{af1a});
on the other hand, if we apply
$A_{\N}(z)$ and $A_{{\N}+1}(e^{\frac{\pi\lambda_n}{\omega_2}})$
to (\ref{rr41}) and demand that similar equations hold for the
function $\psi_{\la_1,\ldots,\la_N}$
(see exact formulae (\ref{rr1}) and (\ref{rr2}) below),
these requirements yield, after an appropriate deformation
of the integration contour, the difference equations
for the Fourier amplitude ${\cal Q}(\bgamma|\blambda)$:
\be\label{rr5a}
i^{N}{\cal Q}(\bgamma+i\omega_1\bfit e_j|\blambda)=
\Big(\frac{\zen^{\omega_1}}{2}\Big)^{-N}\prod_{k=1}^N
\sinh\frac{\textstyle\pi}{\textstyle\omega_2}
(\gamma_j\!-\!\la_k){\cal Q}(\bgamma|\blambda)\,,
\ee
\be\label{rr5b}
i^{N-1}{\cal Q}(\bgamma|\blambda-i\omega_1{\bfit e}'_n)=
\Big(\frac{\zen^{\omega_1}}{2}\Big)^{-N+1}\prod_{j=1}^{N-1}
\sinh\frac{\textstyle\pi}{\textstyle\omega_2}
(\gamma_j\!-\!\la_n){\cal Q}(\bgamma|\blambda)\,,
\ee
where $\{\bfit e'_n\}$ denote the standard basis in $\RR^N$.
It is clear that equations (\ref{rr5a}), (\ref{rr5b}) can be
factorized into the first order equations of Baxter's type.
Assume for the definiteness that
$\frac{\omega_1}{\omega_2}\notin\RR_-$.
Up to a quasiconstant, their solutions can be expressed
in terms of the double sine $S_2(z)$ according to equations
(\ref{ds3}); to eliminate the quasiconstant (which must be set to 1),
we use the dual ($\omega_1\leftrightarrow\omega_2$) counterpart
of (\ref{rr5a}), (\ref{rr5b}) (which arise due to similar reasoning
from (\ref{e1tilde}), (\ref{af1a'})) as well as the requirement
of analyticity with respect to parameters
$\lambda_1,\ldots,\la_{\N}$.

The formal computation sketched above must be matched
by a series of estimates justifying the deformation of the
integration contours. The induction basis is given by a careful study
of the cases, where $N=2,3$. In this way we arrive at the following
result:
\begin{te}\label{main1}
Let $\psi_{\gamma_1,\ldots,\gamma_{{\N}-1}}(x_1,\ldots,x_{{\N}-1})$
be the solution of equations (\ref{e1})-(\ref{af1a'}).
Let $c(\gamma)$ be the rank 1 quantum Harish-Chandra function
defined by (\ref{ch1}). Introduce
\be\label{rr3'}
{\cal Q}(\bgamma|\blambda)=\prod_{j=1}^{N-1}\prod_{n=1}^N
{c}(\gamma_j\!-\!\lambda_n).
\ee
Let $\psi_{\la_1,\ldots,\la_{\N}}$ be the function defined
by the following integral
\be\label{rr4}
\psi_{\la_1,\ldots,\la_{\N}}(x_1,\ldots,x_{\N})=
\int\limits_{\cal C}\mu(\bgamma){\cal Q}(\bgamma|\blambda)
\Psi_{\bgamma,\la_1+\ldots+\la_{\N}}(x_1,\ldots,x_{\N})\,d\bgamma
\ee
where the auxiliary function
$\Psi_{\bgamma,\la_1+\ldots+\la_{\N}}(x_1,\ldots,x_{\N})$ is
defined by (\ref{af2}) and
the contour of integration in the multiple integral is chosen
in such a way that\\
(a). $\mbox{\rm Im}\,\gamma_j>
\mbox{\rm max}_k\,\{\mbox{\rm Im}\,\la_k\}$;\\
(b). The left end of the contour escapes to infinity in the sectors
$$
\frac{1}{2}({\rm arg}\,\omega_1\!+\!{\rm arg}\,\omega_2)+\pi
<{\rm arg}\,{\gamma_j}<{\rm arg}\,\omega_2+\frac{3\pi}{2}\,;
$$
(c). The right end of the contour escapes to infinity in the sectors
$$
{\rm arg}\,\omega_1-\frac{\pi}{2}<{\rm arg}\,{\gamma_j}
<{\rm arg}\,\omega_2+\frac{\pi}{2}\,.
$$
Then the function (\ref{rr4}) is a common eigenfunction for
$N$-particle open RTC. Namely, it satisfies to the following
properties:

(i) $\psi_{\la_1,\ldots,\la_{\N}}$
is an entire function in $\blambda\in\CC^N$;

(ii) $\psi_{\la_1,\ldots,\la_{\N}}$ is the solution to the
following set of equations:
\be\label{rr1}
A_{\N}(z)\psi_{\la_1,\ldots,\la_{\N}}=
\prod_{k=1}^N\Big(z-z^{-1}e^{\frac{2\pi\lambda_k}{\omega_2}}\Big)
\;\psi_{\la_1,\ldots,\la_{\N}}\;,
\ee
\be\label{rr2}
A_{{\N}+1}\Big(e^{\frac{\pi\la_n}{\omega_2}}\Big)
\psi_{\la_1,\ldots,\la_{\N}}\;=\\ =\,q^{-1}(-i\zen^{\omega_1})^{N+1}
\exp\left\{\frac{2\pi}{\omega_2}\sum_{k=1}^N\la_k\right\}\;
e^{-\frac{2\pi{x_{N+1}}}{\omega_2}}
\psi_{\lambda_1,\ldots,\lambda_n-i\omega_1,\ldots,\la_{\N}}\;,
\ee
\be\label{rr1'}
\wt A_{\N}(z)\psi_{\la_1,\ldots,\la_{\N}}=
\prod_{k=1}^N\Big(z-z^{-1}e^{\frac{2\pi\lambda_k}{\omega_1}}\Big)
\;\psi_{\la_1,\ldots,\la_{\N}}\;,
\ee
\be\label{rr2'}
\wt A_{{\N}+1}\Big(e^{\frac{\pi\la_n}{\omega_1}}\Big)
\psi_{\la_1,\ldots,\la_{\N}}\;=\\ =\,q^{-1}(-i\zen^{\omega_2})^{N+1}
\exp\left\{\frac{2\pi}{\omega_1}\sum_{k=1}^N\la_k\right\}\;
e^{-\frac{2\pi{x_{N+1}}}{\omega_1}}
\psi_{\lambda_1,\ldots,\lambda_n-i\omega_2,\ldots,\la_{\N}}\;.
\ee
\end{te}
By inductive application of the formula (\ref{rr4}), starting with
trivial one-particle wave function $\psi_{\gamma_1}(x_1) =
e^{\frac{2\pi i\gamma_1x_1}{\omega_1\omega_2}}$,  we get an explicit
 solution for the $N$-particle system.
\begin{te}
Let $||\gamma_{jk}||_{j,k=1}^N$ be a lower triangular $N\times N$
matrix and let the last row
$(\gamma_{\N1},\ldots,\gamma_{{\N}{\N}})$
be identified with $\!(\la_1,\ldots,\la_{{\N}})\!$.

(i) The solution to (\ref{rr1})-(\ref{rr2'}) can be written
in the form:
\be\label{mb}
\psi_{\la_1,\ldots,\la_N}(x_1,\ldots ,x_{\N})\;=\\
\hspace{-1.6cm}
\int\limits_{{\cal D}_N}\left.\prod_{n=1}^{N-1}\right\{
\!\Big(\prod\limits_{\stackrel{\scriptstyle j,k=1}{j<k}}^n
4\omega_1\omega_2\,\sinh\frac{\pi}{\omega_1}
(\gamma_{nj}\!-\!\gamma_{nk})\cdot
\sinh\frac{\pi}{\omega_2}(\gamma_{nj}\!-\!\gamma_{nk})
\Big)\prod_{j=1}^n\prod_{k=1}^{n+1}
c(\textstyle\gamma_{nj}\!-\!\gamma_{n+1,k})\times
\hspace{-2.4cm}
\\
\exp\left[\left.\frac{\pi i}{\omega_1\omega_2}\left(
\Big(\sum_{m=1}^{N}\gamma_{nm}\Big)^2-
\sum_{k,m=1}^N\gamma_{n+1,k}
\gamma_{nm}\right)\right]\right\}\times
\\
\exp\left[\frac{2\pi i}{\omega_1\omega_2}\sum_{n,m=1}^{N}x_n
\Big(\gamma_{nm}-\gamma_{n-1,m}\Big)\right]
\prod\limits_{\stackrel{\scriptstyle j,k=1}{j\geq k}}^{N-1}
d\gamma_{jk}
\ee
where the integral should be understood as follows.
We integrate from top to bottom of the lower triangular matrix:
first we integrate on  $\gamma_{11}$ over the line
${\rm Im}\,\gamma_{11} >
\max\{{\rm Im}\,\gamma_{21}, {\rm Im}\,\gamma_{22}\};$
then we integrate over the set $(\gamma_{21} ,\gamma_{22})$ along
the lines ${\rm Im}\,\gamma_{2j} >
\max_{m}\{{\rm\,Im}\gamma_ {3m}\}$ and so on. The last integrations
should be performed over the variables
$(\gamma_{{\N}-1,1}\ldots,\gamma_{{\N}-1,{\N}-1})$
along the lines
${\rm Im}\gamma_{{\N}-1,k}>\max_{m}\{{\rm Im}\,\gamma_{{\N},m}\}$.
The asymptotic behaviour of all contours is chosen in the same way
as in the previous theorem.

\smallskip\noindent
(ii) The wave function $\psi_{\blambda}$ has the following asymptotic
behaviour as $x_{j}-x_{k}\to-\infty$ inside the positive Weyl
chamber $P_{+}= \{(x_1,\ldots,x_{\N}); x_1<x_2<...<x_{\N}\}$:
\be\label{hc0}
\psi_{\la_1,\ldots,\la_{\N}}(x_1,\ldots,x_{\N})\sim\sum_{s\in W}
{\rm C}(s\blambda)\,
\exp\Big\{\frac{{\pi}i}{{\omega_1}{\omega_2}}\,
(s\blambda,\bfit x)\Big\}
\ee
where
\be\label{hc1}
{\rm C}(\blambda)\,=\,\prod_{j<k}e^{-\frac{\scriptstyle\pi i}
{\scriptstyle\omega_1\omega_2}\la_j\la_k}{c}(\la_j-\la_k)\,,
\ee
\bigskip\noindent
the sum is over all permutation of $(\lambda_1,\ldots,\lambda_N)$
and $(.\,,.)$ is standard scalar product in $\RR^N$.
\end{te}
We will refer to (\ref{hc1}) as the quantum Harish-Chandra function
for the modular double corresponding to $U_q(\mathfrak{gl}(n))$.
\begin{rem}
In the case of $N=2$ the solution (\ref{mb}) has the following form
\be
\psi_{\la_1,\la_2}(x_1,x_2)\;=\\
\hspace{-0.2cm}
=
e^{\frac{\scriptstyle 2\pi i}{\scriptstyle\omega_1\omega_2}
(\la_1+\la_2)x_2}\int\limits_{{\cal D}_2}
e^{\frac{\scriptstyle\pi i}{\scriptstyle\omega_1\omega_2}
[\gamma_{11}^2-(\la_1+\la_2)\gamma_{_{11}}]}
{c}(\gamma_{_{11}}\!-\!\la_1){c}(\gamma_{_{11}}\!-\!\la_2)\,
e^{\frac{\scriptstyle 2\pi i\gamma_{_{11}}}
{\scriptstyle\omega_1\omega_2}
(x_1-x_2)}\,d\gamma_{_{11}}\;.
\hspace{-0.8cm}
\ee
Changing the integration variable $\gamma_{_{11}}\to\zeta+\lambda_1$
and letting $x=x_1-x_2$ one can obtain (up to simple $GL(1)$ factor)
the function $\psi_{\la_1-\la_2}(x)$ which
coincides with $U_q(\mathfrak{sl}(2))$ solution (\ref{mb1})
for $\e=-1$.
\end{rem}
\begin{cor}
Let the periods ${\omega_1}, {\omega_2}$ be real positive numbers.
Fix the following choice of the coupling constant
$g^{\omega_1}=\frac{q-q^{-1}}{i\omega_1}$ and let $\brho$
be the half-sum of positive roots of $\mathfrak{sl}(N\!,\RR)$
written in standard basis of $\RR^{N}$. After rescaling
$x_{k}\,\to\,\frac{\omega_2}{2\pi}x_k$ and sending
${\omega_2}\to\infty$ one obtains in this limit
$\psi_{\blambda}({\textstyle\frac{\omega_2}{2\pi}}\bfit x)\to
\psi^{(TC)}_{\blambda}(\bfit x)$, where
$\psi^{(TC)}_{\blambda}(\bfit x)$ is a solution of $GL(N,\RR)$ open
Toda chain in the Mellin-Barnes form \cite{KL2}. In terms of the
classical $GL(N,\RR)$ Whittaker functions $W(\bfit x; \blambda)$
\cite{Ha} it can be written in the form
\be\label{we2}
\psi_{\blambda}^{(TC)}(\bfit x)\,=\,
\omega_{1}^{-2i(\blambda,\brho)/\omega_1}
\prod_{j<k}\pi^{-1/2}\Gamma\Big(\frac{\la_j\!-\!\la_k}{i\omega_1}
\!+\!\frac{1}{2}\Big)\;\ W(\bfit x;\blambda),
\ee
In this limit the quantum Harish-Chandra function
(\ref{hc1}) reduces to the standard one:
\be
{\rm C}(\blambda)\,\to\,\omega_{1}^{-2i(\blambda,\brho)/\omega_1}
\prod_{j<k}\Gamma\Big(\frac{\la_j\!-\!\la_k}{i\omega_1}\Big).
\ee
\end{cor}
\begin{rem}
To generalize the solution (\ref{mb1}) to an arbitrary
$N$-particle case one should deal with the Lax operators
\be
L_n^{(\e)}(z)\,=\,\left(\begin{array}{cc}
z-z^{-1}e^{\omega_1p_n}\ \ \ &
g^{\omega_1}e^{-\frac{2\pi{\scriptstyle x_n}}
{\omega_2}+\frac{1+\e}{2}\omega_1p_n}\\
-g^{\omega_1}e^{\frac{2\pi {\scriptstyle x_n}}
{\omega_2}+\frac{1-\e}{2}\omega_1p_n}
& 0
\end{array}\right)\,,
\ee
\be
\wt L_n^{(\e)}(z)\,=\,\left(\begin{array}{cc}
z-z^{-1}e^{\omega_2p_n}\ \ \ &
g^{\omega_2}e^{-\frac{2\pi{\scriptstyle x_n}}
{\omega_1}+\frac{1+\e}{2}\omega_2p_n}\\
-g^{\omega_2}e^{\frac{2\pi{\scriptstyle x_n}}
{\omega_1}+\frac{1-\e}{2}\omega_2p_n}
& 0
\end{array}\right)
\ee
which satisfy the $RTT$ relation with the same twisted $R$-matrix.
Applying the method described above to this model, one obtains the
solution of the same structure as (\ref{mb}) with the additional
factor $-\e$ in front of quadratic form in $\gamma$-variables.
Note that different deformations of the relativistic Toda chain
has been introduced in \cite{Ku} on purely algebraic level.
\end{rem}

\subsection{Periodic $q$-Toda chain}
In this section we will formulate briefly the extension of our
approach to the case of periodic modular $q$-Toda chain.
The details will be published in a separate publication. In the same
way as it was done in the case of the open chain, we can calculate
the action of the transfer matrices $t_{\N}(z)$ and $\wt t_{\N}(z)$
on the auxiliary wave function (\ref{af2}).
One can prove the following result.
\begin{prop}
The action of $t_{\N}(z)$, $\wt t_{\N}(z)$ on the auxiliary wave
function (\ref{af2}) is given by
\be
t_{\N}(z)\Psi_{\bgamma,\varepsilon}=\left(z-z^{-1}
e^{\raise2pt\hbox{$\scriptstyle\frac{2\pi}{\omega_2}\,
\big(\varepsilon\,-\!\sum\limits_{m=1}^{N-1}\gamma_m\big)$}}\right)
\prod_{j=1}^{N-1}\Big(z-z^{-1}e^{\frac{2\pi\gamma_j}{\omega_2}}\Big)
\;\Psi_{\bgamma,\varepsilon}\;+\\
\hspace{-3mm}
e^{\frac{\pi\varepsilon}{\omega_2}}
\sum_{j=1}^{N-1}\Big\{(-i\zen^{\omega_1})^N
\Psi_{\bgamma-i\omega_1\bfit e_j,\varepsilon}+
(i\zen^{\omega_1})^N
\Psi_{\bgamma+i\omega_1\bfit e_j,\varepsilon}\Big\}
\prod_{s\neq j}\frac{z-e^{\frac{2\pi\gamma_s}{\omega_2}}z^{-1}}
{e^{\frac{\pi\gamma_j}{\omega_2}}-e^{\frac{2\pi\gamma_s}{\omega_2}}
e^{-\frac{\pi\gamma_j}{\omega_2}}}\,,
\ee
\be\label{tm1'}
\wt t_{\N}(z)\Psi_{\bgamma,\varepsilon}=\left(z-z^{-1}
e^{\raise2pt\hbox{$\scriptstyle\frac{2\pi}{\omega_1}\,
\big(\varepsilon\,-\!\sum\limits_{m=1}^{N-1}\gamma_m\big)$}}\right)
\prod_{j=1}^{N-1}\Big(z-z^{-1}e^{\frac{2\pi\gamma_j}{\omega_1}}\Big)
\;\Psi_{\bgamma,\varepsilon}\;+\\
\hspace{-3mm}
e^{\frac{\pi\varepsilon}{\omega_1}}
\sum_{j=1}^{N-1}\Big\{(-i\zen^{\omega_2})^N
\Psi_{\bgamma-i\omega_2\bfit e_j,\varepsilon}+
(i\zen^{\omega_2})^N
\Psi_{\bgamma+i\omega_2\bfit e_j,\varepsilon}\Big\}
\prod_{s\neq j}\frac{z-e^{\frac{2\pi\gamma_s}{\omega_1}}z^{-1}}
{e^{\frac{\pi\gamma_j}{\omega_1}}-e^{\frac{2\pi\gamma_s}{\omega_1}}
e^{-\frac{\pi\gamma_j}{\omega_1}}}\;.
\ee
\end{prop}
Using this proposition, one can prove that integral representation
(\ref{rr4}) is still valid for the  wave functions of {\it periodic}
$q$-Toda chain. Namely, the l.h.s. of (\ref{rr4}) should be
considered as the wave function of periodic chain
(i.e. as the common eigenfunction for the operators
$t_{\N}(z)$ and $\wt t_{\N}(z)$) with the following
changes. The Fourier coefficient ${\cal Q}(\bgamma \mid \blambda)$
factorizes now into the product
\be\label{p1}
{\cal Q}(\bgamma \mid \blambda)=\prod_{j=1}^{N-1}
Q(\gamma_j \mid \blambda),
\ee
where $Q(\gamma|\blambda)$ is an entire function with an appropriate
asymptotic behaviour which satisfies the following system of the
mutually dual Baxter equations:
\begin{subequations}\label{p2}
\be\label{p2a}
\hspace{-0.7cm}
\Big(\frac{\zen^{\omega_1}}{2}\Big)^{-N}
\prod_{k=1}^{N}{\sinh{\frac{\pi}{\omega_2}}}(\gamma\!-\!\lambda_k)
Q(\gamma|\blambda)=i^{N}{Q}(\gamma\!+\!i\omega_1|\blambda)+
i^{-N}{Q}(\gamma\!-\!i\omega_1|\blambda),
\ee
\be\label{p2b}
\hspace{-0.7cm}
\Big(\frac{\zen^{\omega_2}}{2}\Big)^{-N}
\prod_{k=1}^{N}{\sinh{\frac{\pi}{\omega_1}}}(\gamma\!-\!\lambda_k)
Q(\gamma|\blambda)=i^{N}Q(\gamma\!+\!i\omega_2|\blambda)+
i^{-N}Q(\gamma\!-\!i\omega_2|\blambda).
\ee
\end{subequations}
(compare with (\ref{rr5a})).
\begin{rem}
In the limit $\omega_2\to\infty$ the equation (\ref{p2a})
goes to the standard Baxter equation for the $N$-particle periodic
Toda chain \cite{Skl1} with Planck constant $\hbar=\omega_1$
provided $g^{\omega_1}=\frac{2\pi}{\omega_2}[1+O(\omega_2^{-1})]$.
\end{rem}

We would like to stress that the proper
asymptotic behaviour of the solution of Baxter equations is fixed
by the condition that the integral (\ref{rr4}) converges.
Together with analyticity condition for the solution this leads to
the quantization conditions of Gutzwiller's type for the eigenvalues
$\blambda$ \cite{Gu}, \cite{PG}, \cite{KL1}.
Note that the Baxter equation (\ref{p2a})
has been obtained in \cite{KT} in the framework of
separation of variables \cite{Skl1}.
A similar {\it system} of Baxter equations (\ref{p2})
appeared for the first time in \cite{FKV}, \cite{Sm2} in the models
different from ours, but with the same type of duality property.

\section*{Acknowledgments}

We are deeply indebted to L. Faddeev, B. Feigin, S. Ruijsenaars,
S. Khoroshkin, and F. Smirnov for stimulating discussions.
We also grateful to T. Sultanov for preparing the pictures.

\smallskip\noindent
The research was partly supported by grants
INTAS 99-01782; RFBR 00-02-16477 (S. Kharchev);
INTAS 97-1312, RFBR 00-02-16530 (D. Lebedev) and by
grant 00-15-96557 for Support of Scientific Schools;
RFBR 99-01-00101 (M. Semenov-Tian-Shansky).

One of us (D.L.) is deeply indebted to Institut des Hautes
\'Etudes Scientifiques where the part of work has been done.

\app{Double sine function}\label{dsf}
We give here a short summary of the properties of the double sines
and related functions. The theory of double sines, double gamma
functions, etc. goes back to the papers of
Barnes \cite{Ba1}, \cite{Ba2}, with more recent additions of
Shintani \cite{Shin} and Kurokawa \cite{Kur}; the closely related
quantum dilogarithms were independently introduced by Faddeev and
Kashaev \cite{FK} in connection with the Quantum Inverse
Scattering Method. See also relevant items on the subject in
\cite{JM}, \cite{Rui3}.

\sapp{Definition and main properties}\label{mp} The basic
properties of double sines listed below are extracted mainly from
\cite{Shin} and  \cite{Kur};   all  main ideas are contained
already in the papers of  Barnes \cite{Ba1}-\cite{Ba3}
\footnote{
In \cite{Ba1, Ba2} Barnes have developed the complete theory of
the so called double gamma functions $\Gamma_2(z|\omega_1,\omega_2)$.
The double sine function appeared for the first time in paper by
Shintani \cite{Shin} as a ratio of two appropriate double
gamma functions.
}
.

\bigskip\noindent
{\bf I. Integral formulae.} Set $\bo=(\omega_1, \omega_2), \
\omega_1, \omega_2 >0$. The double sine   $S_2(z|\bo)$ is defined
by the integral   \cite{Ba1},  \cite{Shin}
\be\label{ds1}
\log S_2(z|\bo)=\int\limits_{{\cal C}_H}
\frac{\sinh(z-\frac{\omega_1+\omega_2}{2})t}{2\sinh\frac{\omega_1t}{2}
\sinh\frac{\omega_2t}{2}}\,\log(-t)\,\frac{dt}{2\pi it}
\ee
in the region
\be\label{ds2}
0<\mbox{Re}\,z<\omega_1\!+\!\omega_2,
\ee
where the   contour ${\cal C}_H$ is drawn on Fig. 1:

\bigskip
\bigskip
\unitlength 1.00mm \linethickness{0.4pt} \hspace{2.3cm}
\begin{picture}(50.00,20.00)
\put(17.00,15.00){\makebox(0,0)[cc]{0}}
\put(20.00,15.00){\makebox(0,0)[cc]{$\bullet$}}
\put(100.00,15.00){\makebox(0,0)[cc]{$+\infty$}}
\linethickness{1pt} \put(20.00,15.00){\line(1,0){70.33}}
\linethickness{0.4pt} \put(40.00,10.00){\line(1,0){50.33}}
\put(90.00,20.00){\line(-1,0){70.00}}
\put(90.00,10.00){\vector(1,0){0.2}}
\put(20.00,10.00){\line(1,0){22.00}}
\put(20.67,10.00){\line(1,0){22.00}}
\put(20.17,15.00){\oval(15.00,10.00)[l]}
\put(40.00,0.00){\makebox(0,0)[cc]{Fig.1 \ \ \
The Hankel contour ${\cal C}_H$}}
\end{picture}

\bigskip\noindent
An equivalent
integral representation  is easily derived from \eqref{ds1}:
\be\label{rep7}
\log S_2(z|\bo)=\frac{\pi i}{2}\,B_{2,2}(z|\bo)\,+\,
\int\limits_{\rs+i0}\frac{e^{zt}}
{(e^{\omega_1t}\!-\!1)(e^{\omega_2t}\!-\!1)}\,\frac{dt}{t},
\ee
where
\be
\label{ds6}
 B_{2,2}(z|\bo)=\frac{z^2}{\omega_1\omega_2}-
\frac{\omega_1+\omega_2}{\omega_1\omega_2}\,z\,+\,
\frac{\omega_1^2+3\omega_1\omega_2+\omega_2^2}{6\omega_1\omega_2}
\ee
and the contour is drawn on Fig. 2:

\bigskip
$$ \unitlength 1.00mm \linethickness{0.4pt}
\begin{picture}(130.00,20.00)
\put(60.00,10.00){\makebox(0,0)[cc]{0}}
\put(60.00,15.00){\makebox(0,0)[cc]{$\scriptstyle\bullet$}}
\put(110.00,15.00){\makebox(0,0)[cc]{${\scriptstyle+}\infty$}}
\put(24.5,15.00){\line(1,0){30.33}}
\put(65.00,15.00){\line(1,0){30.33}}
\put(60,15){\oval(10.00,10.00)[t]}
\put(80.00,14.20){$\scriptstyle\rightarrow$}
\put(40.00,14.20){$\scriptstyle\rightarrow$}
\put(60.7,20.00){\vector(1,0){0.2}}
\put(60.00,0.00){\makebox(0,0)[cc]{Fig.2 \ \ \ Contour $\RR+i0$}}
\end{picture}
$$

\medskip\noindent
Note that
\be\label{ds6'}
B_{2,2}(\omega_1+\omega_2-z|\bo)=B_{2,2}(z|\bo)\,.
\ee

\medskip\noindent
From (\ref{ds1}) one can also derive that
\be\label{dl}
\log S_2(z|\bo)=\int\limits^\infty_0
\left\{\frac{\sinh(z-\frac{\omega_1+\omega_2}{2})t}
{2\sinh\frac{\omega_1t}{2}
\sinh\frac{\omega_2t}{2}}\;-\;\frac{1}{\omega_1\omega_2t}
\,(2z-\omega_1-\omega_2) \right\}\frac{dt}{t}\,.
\ee

{\bf II. Series and product formulae.} Evaluating (\ref{dl}) by
the residue formula, one obtains the series expansions which are
valid in the regions ${\rm Im}\,z>0,$ and ${\rm Im}\,z<0,$
respectively:
\be\label{ds4}
\log S_2(z|\bo)=\frac{\pi i}{2} \,B_{2,2}(z|\bo) +
\sum_{n=1}^\infty\frac{1}{n}\left\{\frac{e^{\frac{\scriptstyle
2\pi inz} {\scriptstyle\omega_1}}}{e^{\frac{\scriptstyle 2\pi
in\omega_2} {\scriptstyle\omega_1}}-1}\,+\,
\frac{e^{\frac{\scriptstyle 2\pi inz}
{\scriptstyle\omega_2}}}{e^{\frac{\scriptstyle 2\pi in\omega_1}
{\scriptstyle\omega_2}}-1}\right\},
\ee
\be\label{ds5}
\log S_2(z|\bo)=\,-\,\frac{\pi i}{2}\,B_{2,2}(z|\bo)+
\sum_{n=1}^\infty\frac{1}{n}\left\{ \frac{e^{-\frac{\scriptstyle
2\pi inz} {\scriptstyle\omega_1}}}{e^{-\frac{\scriptstyle 2\pi
in\omega_2} {\scriptstyle\omega_1}}-1}\,+\,
\frac{e^{-\frac{\scriptstyle 2\pi inz}
{\scriptstyle\omega_2}}}{e^{-\frac{\scriptstyle 2\pi in\omega_1}
{\scriptstyle\omega_2}}-1}\right\}.
\ee
By an appropriate deformation of the contour ${\cal C}_H$ the
double sine (\ref{ds1}) can be extended to all complex values of
$\omega_1, \ \omega_2 $, provided that
$\frac{\omega_1}{\omega_2}\notin\RR_-$ \cite{Ba1}. Namely, the
contour is in general along the bisector of the smallest angle
between $-{\rm arg}\,\omega_1$ and $-{\rm arg}\,\omega_2$
enclosing the origin but no other poles of the integrand in
(\ref{ds1}) (the contour $\RR+i0$ should be rotated similarly). As
a corollary, we obtain the product expansions which are valid when
${\rm Im}\frac{\omega_1}{\omega_2}>0$):
\be\label{sh1}
S_2(z|\bo)\,=\,e^{\frac{\scriptstyle\pi i}{2}B_{2,2}(z|\bo)}
\frac{\prod\limits_{m=0}^\infty\left(1-q^{2m}
e^{\frac{\scriptstyle 2\pi iz}{\scriptstyle\omega_2}}\right)}
{\prod\limits_{m=1}^\infty\left(1-\wt q^{\;-2m}
e^{\frac{\scriptstyle 2\pi iz}{\scriptstyle\omega_1}}\right)}
\;=\\
e^{-\frac{\scriptstyle\pi i}{2}B_{2,2}(z|\bo)}
\frac{\prod\limits_{m=0}^\infty\left(1-\wt q^{\;-2m}
e^{-\frac{\scriptstyle 2\pi iz}{\scriptstyle\omega_1}}\right)}
{\prod\limits_{m=1}^\infty\left(1-q^{2m} e^{-\frac{\scriptstyle
2\pi iz}{\scriptstyle\omega_2}}\right)}\ ,
\ee
where
$$
q=e^{\pi i\frac{\omega_1}{\omega_2}}, \ \ \ \ \
\wt q=e^{\pi i\frac{\omega_2}{\omega_1}}.
$$
The equality of the two expressions is due to the modular
transformation law for the theta function
$\theta_1(z|\frac{\omega_1}{\omega_2})$.

\medskip\noindent
{\bf III. Functional relations.} The function $S_2(z|\bo)$
satisfies the difference equations
\begin{subequations}\label{ds3}
\be\label{ds3a}
\frac{S_2(z+\omega_1|\bo)}{S_2(z|\bo)}=
\frac{1}{2\sin\frac{\textstyle\pi z}{\textstyle\omega_2}}\,,
\ee
\be\label{ds3b}
\frac{S_2(z+\omega_2|\bo)}{S_2(z|\bo)}=
\frac{1}{2\sin\frac{\textstyle\pi z}{\textstyle\omega_1}}\,.
\ee
\end{subequations}
Moreover,
\be\label{ds3'}
S_2(z|\bo)S_2(-z|\bo)=-4\sin\frac{\pi z}{\omega_1}\sin\frac{\pi z}
{\omega_2}\,,\\
S_2(z|\bo)S_2(\omega_1+\omega_2-z|\bo)=1\,.
\ee

\bigskip\noindent
{\bf IV. Poles and zeroes.} The zeroes and poles of $S_2(z|\bo)$
are as follows:
\be\label{zp}
\mbox{poles at}\ \ z=n_1\omega_1+n_2\omega_2, \; \;  n_1,\
n_2\geq 1,\\ \mbox{zeros at}\ \ z =n_1\omega_1+n_2\omega_2, \; \;
n_1,n_2\leq 0.
\ee
Moreover,
\be\label{ds7}
\lim_{z\to 0}z^{-1}S_2(z|\bo)=\frac{2\pi}{\sqrt{\omega_1\omega_2}}\,.
\ee
Hence, from (\ref{ds3}) and (\ref{ds7}),
\be\label{om}
S_2(\omega_1|\bo)\  = \  \sqrt{\frac{\omega_2}{\omega_1}}\,, \; \;
\; S_2(\omega_2|\bo)\  = \  \sqrt{\frac{\omega_1}{\omega_2}}\,.
\ee
Using (\ref{ds3}),
(\ref{ds7}), one can calculate the residues of $S_2(z|\bo)$ and
$S_2^{-1}(z|\bo)$ at the corresponding points (\ref{zp}):
\begin{subequations}\label{ds9}
\be\label{ds9a}
\hspace{-0.3cm}
\lim_{z\to 0}zS_2(z\!+\!n_1\omega_1\!+\!n_2\omega_2|\bo)=
\frac{\sqrt{\omega_1\omega_2}}{2\pi}\;
\frac{(-1)^{n_1n_2}}{\prod\limits_{k=1}^{n_1-1}
2\sin\frac{\textstyle\pi k\omega_1}{\textstyle\omega_2}
\prod\limits_{m=1}^{n_2-1} 2\sin\frac{\textstyle\pi m\omega_2}
{\textstyle\omega_1}},
\ee
\be\label{ds9b}
\hspace{-0.3cm}
\lim_{z\to 0}zS^{-1}_2(z\!-\!n_1\omega_1\!-\!n_2\omega_2|\bo)=
\frac{\sqrt{\omega_1\omega_2}}{2\pi}\;
\frac{(-1)^{n_1n_2+n_1+n_2}}{\prod\limits_{k=1}^{n_1}
2\sin\frac{\textstyle\pi k\omega_1}{\textstyle\omega_2}
\prod\limits_{m=1}^{n_2} 2\sin\frac{\textstyle\pi m\omega_2}
{\textstyle\omega_1}}.
\ee
\end{subequations}

\bigskip\noindent
{\bf V. Asymptotics.}
Assuming that ${\rm Im}\,\frac{\omega_1}{\omega_2}>0$, we have
\begin{equation*}
S_2(z|\bo)\; \raise-5pt
\hbox{$\stackrel{-\!\!-\!\!\!\longrightarrow}{\scriptstyle
z\to\infty}$} \;\left\{
\begin{array}{ll}
e^{\frac{\pi i}{2}B_{2,2}(z|\bo)}, & {\rm arg}\,\omega_1<{\rm
arg}\,z
<{\rm arg}\,\omega_2+\pi,\\
e^{-\frac{\pi i}{2}B_{2,2}(z|\bo)}, & {\rm arg}\,
\omega_1-\pi<{\rm arg}\,z<{\rm arg}\,\omega_2, \\
\frac{\textstyle e^{-\frac{\pi i}{2}B_{2,2}(z|\bo)}}
{\textstyle\prod\limits_{m=1}^\infty\left(1-q^{2m}
e^{-\frac{\scriptstyle 2\pi iz}{\scriptstyle\omega_2}}\right)}, &
{\rm arg}\,\omega_2<{\rm arg}\,z<{\rm arg}\,\omega_1,\\
e^{\frac{\pi i}{2}B_{2,2}(z|\bo)}
\prod\limits_{m=0}^\infty\left(1-q^{2m}
e^{\frac{\scriptstyle 2\pi iz}{\scriptstyle\omega_2}}\right), &
{\rm arg}\,\omega_2\!+\!\pi<{\rm arg}\,z
<{\rm arg}\,\omega_1\!+\!\pi.
\end{array}\right.
\end{equation*}

\bigskip\noindent
{\bf VI. Complex conjugation.}
\be\label{ds10}
\ov{S_2(z|\bo)}=S_2(\ov z\,|\ov\bo)\,.
\ee

\sapp{Related functions} It is convenient to introduce the
function ${\cal S}(z|\bo)$ as follows:
\be\label{cs}
S_2(z|\bo)=e^{\frac{\pi i}{2}B_{2,2}(z|\bo)}{\cal S}(z|\bo),
\ee
where according to (\ref{rep7})
\be\label{srep1}
\log {\cal S}(z|\bo)\,=\,\int\limits_{\rs+i0}\frac{e^{zt}}
{(e^{\omega_1t}\!-\!1)(e^{\omega_2t}\!-\!1)}\,\frac{dt}{t}.
\ee
For complex periods with ${\rm Im}\,\frac{\omega_1}{\omega_2}>0$ we
get
\be\label{srep2}
\hspace{-0.7cm}
{\cal S}(z|\bo)\,=\,
\frac{\prod\limits_{m=0}^\infty\left(1-q^{2m}
e^{\frac{\scriptstyle 2\pi iz}{\scriptstyle\omega_2}}\right)}
{\prod\limits_{m=1}^\infty\left(1-\wt q^{\;-2m}
e^{\frac{\scriptstyle 2\pi iz}{\scriptstyle\omega_1}}\right)}
\;=\;e^{-\pi iB_{2,2}(z|\bo)}\;
\frac{\prod\limits_{m=0}^\infty\left(1-\wt q^{\;-2m}
e^{-\frac{\scriptstyle 2\pi iz}{\scriptstyle\omega_1}}\right)}
{\prod\limits_{m=1}^\infty\left(1-q^{2m} e^{-\frac{\scriptstyle
2\pi iz}{\scriptstyle\omega_2}}\right)}.
\ee
Evidently, ${\cal S}(z|\bo)$ satisfies the difference equations
\begin{subequations}\label{cs1}
\be\label{cs1a}
\frac{{\cal S}(z+\omega_1|\bo)}{{\cal S}(z|\bo)}=
\frac{1}{1-e^{\frac{2\pi iz}{\omega_2}}},
\ee
\be\label{cs1b}
\frac{{\cal S}(z+\omega_2|\bo)}{{\cal S}(z|\bo)}=
\frac{1}{1-e^{\frac{2\pi iz}{\omega_1}}}
\ee
\end{subequations}
and obeys the condition
\be\label{ds11}
\ov{S(z|\bo)}=S^{-1}(\omega_1+\omega_2-\ov z\,|\bo)
\ee
when $\omega_1,\,\omega_2$ are real or $\ov\omega_1=\omega_2$.

Let us finally mention that  the {\it quantum dilogarithm}
$e_\bo(z)$ defined in \cite{Fad2}, \cite{FK} is related to
${\cal S}(z|\bo)$ via
\be\label{xyz}
e_\bo(z)={\cal S}(\frac{\omega_1+\omega_2}{2}-iz|\bo)\,,
\ee
while the {\it hyperbolic gamma function} $G(z|\bo)$ introduced in
\cite{Rui2} is
\be\label{es11}
G(z|\bo)=S_2(\frac{\omega_1\!+\!\omega_2}{2}\,+iz|\bo)\,.
\ee

\sapp{Fourier transform}\label{ftr}
The function ${\cal S}(it|\bo)$ has the following
asymptotics:
\be\label{as1}
{\cal S}(it|\bo)\;
\raise-5pt
\hbox{$\stackrel{-\!\!-\!\!\!\longrightarrow}
{\scriptstyle t\to\infty}$}\;\left\{
\begin{array}{ll}
1, &
{\rm arg}\,\omega_1-\frac{\pi}{2}<{\rm arg}\,t
<{\rm arg}\,\omega_2+\frac{\pi}{2}\\
e^{-\pi iB_{2,2}(it|\bo)}, &
{\rm arg}\,\omega_1+\frac{\pi}{2}<{\rm arg}\,t<{\rm arg}\,\omega_2
+\frac{3\pi}{2}
\end{array}\right.
\ee
(we are not interested in behavior in two remaining sectors).
Each sector is divided into two subsectors $(+)$ (resp. $(-)$) where
the exponential $e^{\frac{\pi it^2}{\omega_1\omega_2}}$ (resp.
$e^{-\frac{\pi it^2}{\omega_1\omega_2}}$)
is rapidly decreasing.
To be more precise, the sectors $(+)$ are determined by inequalities
\begin{subequations}\label{s+}
\be\label{s+1}
\frac{1}{2}({\rm arg}\,\omega_1\!+\!{\rm arg}\,\omega_2)+\pi
<{\rm arg}\,t<{\rm arg}\,\omega_2+\frac{3\pi}{2}\,,\\
\ee
\be\label{s+2}
\frac{1}{2}({\rm arg}\,\omega_1\!+\!{\rm arg}\,\omega_2)
<{\rm arg}\,t<{\rm arg}\,\omega_2+\frac{\pi}{2}\,,
\ee
\end{subequations}
while the sectors $(-)$ are determined by inequalities
\begin{subequations}\label{s-}
\be\label{s-1}
{\rm arg}\,\omega_1+\frac{\pi}{2}<{\rm arg}\,t
<\frac{1}{2}({\rm arg}\,\omega_1\!+\!{\rm arg}\,\omega_2)+\pi\,,
\ee
\be\label{s-2}
{\rm arg}\,\omega_1-\frac{\pi}{2}<{\rm arg}\,t
<\frac{1}{2}({\rm arg}\,\omega_1\!+\!{\rm arg}\,\omega_2)\,.
\ee
\end{subequations}

(see Fig.3 for the typical complex periods
$\omega_1$ and $\omega_2$).

\mathsurround=0pt
$$
\special{em:linewidth 1.0pt}
\unitlength 1.00mm
\linethickness{1.0pt}
\begin{picture}(101.67,110.50)
\put(65.75,104.20){\makebox(0,0)[cc]
{\colorbox{g3}{\textcolor{g3}{\tiny ......................}}}}
\put(65.95,103.32){\makebox(0,0)[cc]
{\colorbox{g3}{\textcolor{g3}{\tiny ....................}}}}
\put(65.65,102.40){\makebox(0,0)[cc]
{\colorbox{g3}{\textcolor{g3}{\tiny ....................}}}}
\put(65.60,100.20){\makebox(0,0)[cc]
{\colorbox{g3}{\textcolor{g3}{\tiny ...................}}}}
\put(65.57,98.80){\makebox(0,0)[cc]
{\colorbox{g3}{\textcolor{g3}{\tiny ..................}}}}
\put(65.20,97.55){\makebox(0,0)[cc]
{\colorbox{g3}{\textcolor{g3}{\tiny ..................}}}}
\put(65.15,96.00){\makebox(0,0)[cc]
{\colorbox{g3}{\textcolor{g3}{\tiny ................}}}}
\put(65.00,94.10){\makebox(0,0)[cc]
{\colorbox{g3}{\textcolor{g3}{\tiny ...............}}}}
\put(64.85,93.40){\makebox(0,0)[cc]
{\colorbox{g3}{\textcolor{g3}{\tiny ...............}}}}
\put(64.87,92.67){\makebox(0,0)[cc]
{\colorbox{g3}{\textcolor{g3}{\tiny ..............}}}}
\put(64.75,91.00){\makebox(0,0)[cc]
{\colorbox{g3}{\textcolor{g3}{\tiny .............}}}}
\put(64.55,89.60){\makebox(0,0)[cc]
{\colorbox{g3}{\textcolor{g3}{\tiny ............}}}}
\put(64.40,88.00){\makebox(0,0)[cc]
{\colorbox{g3}{\textcolor{g3}{\tiny ...........}}}}
\put(64.20,86.00){\makebox(0,0)[cc]
{\colorbox{g3}{\textcolor{g3}{\tiny ..........}}}}
\put(64.00,84.33){\makebox(0,0)[cc]
{\colorbox{g3}{\textcolor{g3}{\tiny .........}}}}
\put(63.77,82.67){\makebox(0,0)[cc]
{\colorbox{g3}{\textcolor{g3}{\tiny ........}}}}
\put(63.57,80.67){\makebox(0,0)[cc]
{\colorbox{g3}{\textcolor{g3}{\tiny .......}}}}
\put(63.37,79.77){\makebox(0,0)[cc]
{\colorbox{g3}{\textcolor{g3}{\tiny ......}}}}
\put(63.25,77.60){\makebox(0,0)[cc]
{\colorbox{g3}{\textcolor{g3}{\tiny .....}}}}
\put(63.10,75.70){\makebox(0,0)[cc]
{\colorbox{g3}{\textcolor{g3}{\tiny ....}}}}
\put(62.70,73.82){\makebox(0,0)[cc]
{\colorbox{g3}{\textcolor{g3}{\tiny ...}}}}
\put(62.65,71.50){\makebox(0,0)[cc]
{\colorbox{g3}{\textcolor{g3}{\tiny .}}}}
\put(62.47,69.60){\makebox(0,0)[cc]
{\colorbox{g3}{\textcolor{g3}{\tiny $\!$}}}}
\put(62.15,69.00){\makebox(0,0)[cc]
{\colorbox{g3}{\textcolor{g3}{\tiny $\!\!$}}}}
\put(61.90,67.00){\makebox(0,0)[cc]
{\colorbox{g3}{\textcolor{g3}{\tiny $\!\!\!$}}}}
\put(58.73,31.20){\makebox(0,0)[cc]
{\colorbox{g3}{\textcolor{g3}{\tiny
...........$\!$........}}}}
\put(59.00,32.45){\makebox(0,0)[cc]
{\colorbox{g3}{\textcolor{g3}{\tiny
 ..................}}}}
\put(58.85,34.00){\makebox(0,0)[cc]
{\colorbox{g3}{\textcolor{g3}{\tiny
 ........$\!$.........}}}}
\put(59.00,35.90){\makebox(0,0)[cc]
{\colorbox{g3}{\textcolor{g3}{\tiny ...............}}}}
\put(59.15,36.60){\makebox(0,0)[cc]
{\colorbox{g3}{\textcolor{g3}{\tiny ...............}}}}
\put(59.13,37.33){\makebox(0,0)[cc]
{\colorbox{g3}{\textcolor{g3}{\tiny ..............}}}}
\put(59.25,39.00){\makebox(0,0)[cc]
{\colorbox{g3}{\textcolor{g3}{\tiny .............}}}}
\put(59.45,40.40){\makebox(0,0)[cc]
{\colorbox{g3}{\textcolor{g3}{\tiny ............}}}}
\put(59.60,42.00){\makebox(0,0)[cc]
{\colorbox{g3}{\textcolor{g3}{\tiny ...........}}}}
\put(59.80,44.00){\makebox(0,0)[cc]
{\colorbox{g3}{\textcolor{g3}{\tiny ..........}}}}
\put(60.00,45.67){\makebox(0,0)[cc]
{\colorbox{g3}{\textcolor{g3}{\tiny .........}}}}
\put(60.23,47.33){\makebox(0,0)[cc]
{\colorbox{g3}{\textcolor{g3}{\tiny ........}}}}
\put(60.43,49.33){\makebox(0,0)[cc]
{\colorbox{g3}{\textcolor{g3}{\tiny .......}}}}
\put(60.63,50.23){\makebox(0,0)[cc]
{\colorbox{g3}{\textcolor{g3}{\tiny ......}}}}
\put(60.75,52.40){\makebox(0,0)[cc]
{\colorbox{g3}{\textcolor{g3}{\tiny .....}}}}
\put(60.90,54.30){\makebox(0,0)[cc]
{\colorbox{g3}{\textcolor{g3}{\tiny ....}}}}
\put(61.20,56.18){\makebox(0,0)[cc]
{\colorbox{g3}{\textcolor{g3}{\tiny ...}}}}
\put(61.32,57.50){\makebox(0,0)[cc]
{\colorbox{g3}{\textcolor{g3}{\tiny ..}}}}
\put(61.42,58.90){\makebox(0,0)[cc]
{\colorbox{g3}{\textcolor{g3}{\tiny .}}}}
\put(61.55,60.40){\makebox(0,0)[cc]
{\colorbox{g3}{\textcolor{g3}{\tiny $\!$}}}} {\boldmath
\put(38.79,96.17){\makebox(0,0)[cc]{$\scriptstyle\omega_1$}}
\put(101.67,65.00){\makebox(0,0)[cc]
{\large $\scriptstyle{\cal S}(it|\bo)\to 1$}}
\put(40.67,87.80){\makebox(0,0)[cc]{$\scriptstyle\omega_2$}}
\put(13.50,65.50){\makebox(0,0)[cc]
{\large $\scriptstyle e^{-\pi iB_{2,2}(it|\bo)}
\gets {\cal S}(it|\bo)$}} }
\put(49.50,105.00){\makebox(0,0)[cc]
{$\scriptstyle{\rm arg}\,\omega_1+\frac{\pi}{2}$}}
\put(70.00,110.50){\makebox(0,0)[cc]
{$\scriptstyle\frac{1}{2}({\rm arg}
\,\omega_1+{\rm arg}\,\omega_2+\pi)$}}
\put(31.00,71.67){\makebox(0,0)[cc]
{$\scriptstyle\frac{1}{2}({\rm arg}
\,\omega_1+{\rm arg}\,\omega_2)+\pi$}}
\put(83.00,105.00){\makebox(0,0)[cc]
{$\scriptstyle{\rm arg}\,\omega_2+\frac{\pi}{2}$}}
\put(43.33,28.33){\makebox(0,0)[cc]
{$\scriptstyle{\rm arg}\,\omega_2-\frac{\pi}{2}$}}
\put(73.67,28.33){\makebox(0,0)[cc]
{$\scriptstyle{\rm arg}\,\omega_1-\frac{\pi}{2}$}}
\put(89.50,58.00){\makebox(0,0)[cc]
{$\scriptstyle\frac{1}{2}
({\rm arg}\,\omega_1+{\rm arg}\,\omega_2)$}}
\emline{70.00}{75.00}{1}{76.00}{75.00}{2}
\emline{73.00}{78.00}{3}{73.00}{72.00}{4}
\emline{46.33}{75.00}{5}{52.33}{75.00}{6}
\emline{46.33}{55.00}{7}{52.33}{55.00}{8}
\emline{49.33}{58.00}{9}{49.33}{52.00}{10}
\emline{69.67}{55.00}{11}{75.67}{55.00}{12}
\emline{70.02}{75.02}{13}{76.02}{75.02}{14}
\emline{73.02}{78.02}{15}{73.02}{72.02}{16}
\emline{46.35}{55.02}{17}{52.35}{55.02}{18}
\emline{49.35}{58.02}{19}{49.35}{52.02}{20}
\emline{46.33}{75.08}{21}{52.33}{75.08}{22}
\emline{69.67}{55.02}{23}{75.67}{55.02}{24}
{\special{em:linewidth 0.4pt}\linethickness{0.4pt}
\emline{62.00}{30.00}{25}{62.00}{105.00}{26}
\emline{33.00}{65.00}{137}{90.33}{65.00}{138}
}
\emline{50.34}{30.04}{27}{75.32}{104.79}{28}
\emline{56.68}{104.79}{29}{66.81}{30.04}{30}
\put(38.65,93.78){\vector(4,1){0.2}}
\emline{30.99}{92.66}{31}{38.65}{93.69}{32}
\emline{30.99}{92.76}{33}{38.65}{93.79}{34}
\put(42.02,89.67){\vector(4,-1){0.2}}
\emline{30.99}{92.66}{35}{42.02}{89.67}{36}
\emline{30.99}{92.57}{37}{42.02}{89.58}{38}
\emline{30.99}{86.03}{39}{30.99}{99.86}{40}
\emline{42.86}{92.66}{41}{19.59}{92.66}{42}
\emline{56.64}{104.83}{43}{66.77}{30.08}{44}
\emline{56.60}{104.87}{45}{66.73}{30.12}{46}
\emline{50.42}{30.00}{47}{75.40}{104.75}{48}
\emline{50.38}{30.04}{49}{75.36}{104.79}{50}
\emline{56.07}{29.36}{51}{69.24}{107.96}{52}
\emline{56.59}{104.89}{53}{66.72}{30.14}{54}
\emline{56.55}{104.93}{55}{66.68}{30.18}{56}
\emline{56.51}{104.97}{57}{66.64}{30.22}{58}
\emline{50.30}{30.04}{59}{75.28}{104.79}{60}
\emline{50.38}{30.00}{61}{75.36}{104.75}{62}
\emline{56.03}{29.40}{63}{69.20}{108.00}{64}
\emline{56.11}{29.32}{65}{69.28}{107.92}{66}
\emline{34.92}{69.49}{67}{87.91}{60.78}{68}
\emline{34.92}{69.46}{69}{87.91}{60.75}{70}
\emline{34.95}{69.43}{71}{87.94}{60.72}{72}
\emline{34.95}{69.40}{73}{87.94}{60.69}{74}
\emline{56.08}{29.32}{75}{69.25}{107.92}{76}
\emline{31.00}{105.94}{77}{33.70}{105.66}{78}
\emline{33.70}{105.66}{79}{36.29}{104.84}{80}
\emline{36.29}{104.84}{81}{38.66}{103.50}{82}
\emline{38.66}{103.50}{83}{40.71}{101.72}{84}
\emline{40.71}{101.72}{85}{42.35}{99.55}{86}
\emline{42.35}{99.55}{87}{43.51}{97.10}{88}
\emline{43.51}{97.10}{89}{44.15}{94.45}{90}
\emline{44.15}{94.45}{91}{44.24}{91.74}{92}
\emline{44.24}{91.74}{93}{43.77}{89.06}{94}
\emline{43.77}{89.06}{95}{42.77}{86.54}{96}
\emline{42.77}{86.54}{97}{41.27}{84.27}{98}
\emline{41.27}{84.27}{99}{39.35}{82.35}{100}
\emline{39.35}{82.35}{101}{37.07}{80.87}{102}
\emline{37.07}{80.87}{103}{34.54}{79.88}{104}
\emline{34.54}{79.88}{105}{31.86}{79.42}{106}
\emline{31.86}{79.42}{107}{29.15}{79.53}{108}
\emline{29.15}{79.53}{109}{26.51}{80.18}{110}
\emline{26.51}{80.18}{111}{24.06}{81.35}{112}
\emline{24.06}{81.35}{113}{21.90}{83.00}{114}
\emline{21.90}{83.00}{115}{20.13}{85.06}{116}
\emline{20.13}{85.06}{117}{18.80}{87.43}{118}
\emline{18.80}{87.43}{119}{17.99}{90.03}{120}
\emline{17.99}{90.03}{121}{17.73}{92.73}{122}
\emline{17.73}{92.73}{123}{18.02}{95.43}{124}
\emline{18.02}{95.43}{125}{18.86}{98.02}{126}
\emline{18.86}{98.02}{127}{20.20}{100.38}{128}
\emline{20.20}{100.38}{129}{22.00}{102.42}{130}
\emline{22.00}{102.42}{131}{24.17}{104.05}{132}
\emline{24.17}{104.05}{133}{26.63}{105.20}{134}
\emline{26.63}{105.20}{135}{31.00}{105.94}{136}
\end{picture}
$$
\mathsurround=2pt
\begin{center}
\vspace{-2cm}

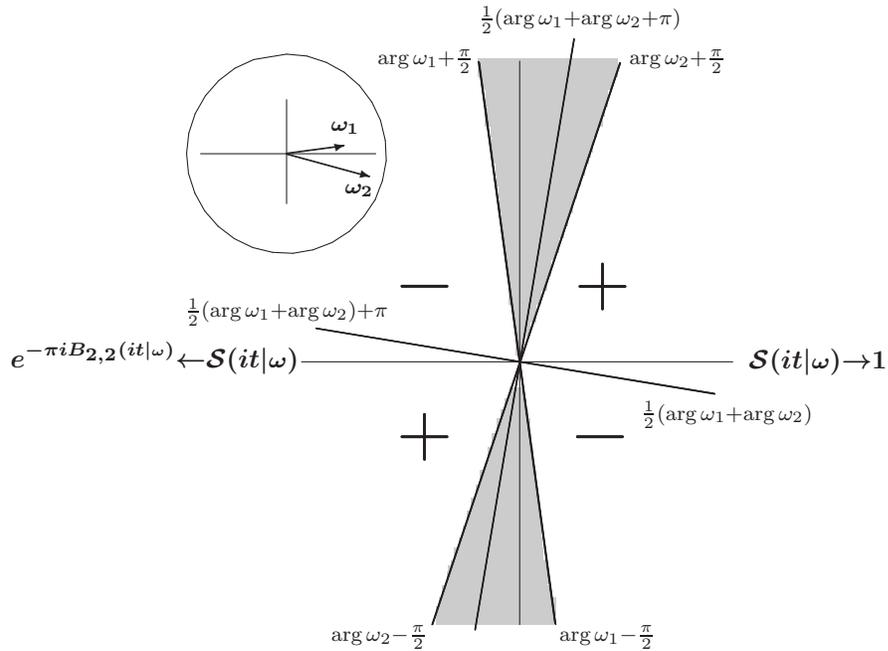
{Fig. 3. }
\end{center}

There exist remarkable integral formulae with the double sine type
functions \cite{FKV, PT}. In particular, there are the following
Fourier transformation formulae \cite{FKV}
\footnote{
Actually, any three formulae in (\ref{ft1}) are consequences of the
fourth one. Nevertheless, we list all of them for convenience.
}
:
\begin{subequations}\label{ft1}
\be
\label{ft1a}
\int\limits_{\Gamma}
{\cal S}(it\!+\!\omega_1\!+\omega_2\!-\!a)\,
e^{\frac{2\pi izt}{\omega_1\omega_2}}dt=
\sqrt{\omega_1\omega_2}\;e^{-\frac{\pi i}{2}B_{2,2}(0)}\;
{\cal S}^{-1}(-iz)\;e^{\frac{2\pi za}{\omega_1\omega_2}}\\
\ee
\be\label{ft2b}
\int\limits_{{\rm L}}{\cal S}^{-1}(it+a)\,
e^{\frac{2\pi izt}{\omega_1\omega_2}}dt=
\sqrt{\omega_1\omega_2}\;e^{\frac{\pi i}{2}B_{2,2}(0)}\;
{\cal S}(\omega_1+\omega_2+iz)\;
e^{-\frac{2\pi za}{\omega_1\omega_2}}
\ee
\be\label{ft2a}
\int\limits_{\Gamma'}
{\cal S}(-it\!+\!\omega_1\!+\!\omega_2\!-\!a)\,
e^{\frac{2\pi izt}{\omega_1\omega_2}}dt=
\sqrt{\omega_1\omega_2}\;e^{-\frac{\pi i}{2}B_{2,2}(0)}\;
{\cal S}^{-1}(iz)\;e^{-\frac{2\pi za}{\omega_1\omega_2}}\\
\ee
\be\label{ft1b}
\int\limits_{{\rm L}'}{\cal S}^{-1}(-it+a)\,
e^{\frac{2\pi izt}{\omega_1\omega_2}}dt=
\sqrt{\omega_1\omega_2}\;e^{\frac{\pi i}{2}B_{2,2}(0)}\;
{\cal S}(\omega_1+\omega_2-iz)\;
e^{\frac{2\pi za}{\omega_1\omega_2}}
\ee
\end{subequations}

\smallskip\noindent
The notations here are as follows.
The contours $\Gamma$ and ${\rm L}'$ are above the poles
\be
\hspace{1cm}
t_{n_1,n_2}=-i(a+n_1\omega_1+n_2\omega_2)\hspace{2cm}(n_1,n_2\geq 0)
\ee
of the integrands in (\ref{ft1a}) and (\ref{ft1b}) while the
contours ${\rm L}$ and $\Gamma'$ are below the poles
\be
\hspace{1.2cm}
t'_{n_1,n_2}=i(a+n_1\omega_1+n_2\omega_2)\hspace{2cm}(n_1,n_2\geq 0)
\ee
of the integrands in (\ref{ft2b}) and (\ref{ft2a}).\\
Further, the contours $\Gamma$ and ${\rm L}$ are beginning
in subsectors (\ref{s+1}) and (\ref{s-1}) respectively,
but may lie in the whole sector $[{\rm arg}\,\omega_1-\frac{\pi}{2}
\,,{\rm arg}\,\omega_2+\frac{\pi}{2}]$,
while the contours $\Gamma'$ and ${\rm L}'$
may lie in the whole sector $[{\rm arg}\,\omega_1+\frac{\pi}{2}\,,
{\rm arg}\,\omega_2+\frac{3\pi}{2}]$, but are ending in subsectors
(\ref{s+2}) and (\ref{s-2}), respectively.

\bigskip\noindent
Provided such a description, the formulae (\ref{ft1a}) and
(\ref{ft2b}) hold in the region
\be
\label{restr1}
{\rm arg}\,z\notin \Big[{\rm arg}\,\omega_2-\frac{\pi}{2},
{\rm arg}\,\omega_1-\frac{\pi}{2}\Big]\,,
\ee
while in (\ref{ft2a}) and (\ref{ft1b})
\be
\label{restr2}
{\rm arg}\,z\notin \Big[{\rm arg}\,\omega_2+\frac{\pi}{2},
{\rm arg}\,\omega_1+\frac{\pi}{2}\Big]\,.
\ee

\newcommand{\CMP}[3]{{\it Comm. Math. Phys. }{\bf #1} (#2) #3}
\newcommand{\LMP}[3]{{\it Lett. Math. Phys. }{\bf #1} (#2) #3}
\newcommand{\IMP}[3]{{\it Int. J. Mod. Phys. }{\bf A#1} (#2) #3}
\newcommand{\NP}[3]{{\it Nucl. Phys. }{\bf B#1} (#2) #3}
\newcommand{\PL}[3]{{\it Phys. Lett. }{\bf B#1} (#2) #3}
\newcommand{\MPL}[3]{{\it Mod. Phys. Lett. }{\bf A#1} (#2) #3}
\newcommand{\PRL}[3]{{\it Phys. Rev. Lett. }{\bf #1} (#2) #3}
\newcommand{\AP}[3]{{\it Ann. Phys. (N.Y.) }{\bf #1} (#2) #3}
\newcommand{\LMJ}[3]{{\it Leningrad Math. J. }{\bf #1} (#2) #3}
\newcommand{\FAA}[3]{{\it Funct. Anal. Appl. }{\bf #1} (#2) #3}
\newcommand{\PTPS}[3]{{\it Progr. Theor. Phys. Suppl. }
{\bf #1} (#2) #3}
\newcommand{\LMN}[3]{{\it Lecture Notes in Mathematics }
{\bf #1} (#2) #2}

\end{document}